\documentclass[%
 reprint,
 amsmath,amssymb,
 aps,
 prd,
 showkeys,
]{revtex4-2}
\usepackage{graphicx}
\usepackage{dcolumn}
\usepackage{bm}
\usepackage{color}

\newcommand{\be}{\begin{eqnarray}}
\newcommand{\ee}{\end{eqnarray}}
\newcommand{\bea}{\begin{eqnarray}}
\newcommand{\eea}{\end{eqnarray}}

\begin{document}

\preprint{APS/123-QED}

\title{Polarization Signatures of Magnetic Reconnection Bubbles near Black Holes}

\author{Yu-Xiang Huang$^{1}$, En-Wei Liang$^{1*}$}

\affiliation{
$^1$Guangxi Key Laboratory for Relativistic Astrophysics, School of Physical Science and Technology, Guangxi University, Nanning 530004, People's Republic of China\\
}

\email{$^1$ yxhuangphys@126.com \\  $^1$ lew@gxu.edu.cn (Corresponding author) }

\date{\today}
\begin{abstract}
Motivated by the observed horizon-scale polarization variability in multi-epoch Event Horizon Telescope (EHT) data, we develop a time-dependent general relativistic polarized ray-tracing framework to study how localized magnetic reconnection near a black hole imprints itself on synchrotron polarization. We model the reconnection phenomenologically as a transient magnetic bubble. While keeping the underlying spacetime, emission prescription, and polarization transport scheme unchanged, the bubble weakens the local ordered field, reorganizes its geometry, induces a conservative local kinematic response, and relaxes through a finite magnetic-memory prescription. We explore how the key bubble parameters control the recovery of the magnetic-field strength, the spatial spreading of the disturbed region, the persistence of the post-reconnection imprint, and the resulting polarization signatures. Perturbations dominated by different magnetic-field components produce distinct polarized morphologies, EVPA evolution, and Stokes-$(Q, U)$ trajectories: azimuthal perturbations favor outward EVPA shifts, radial perturbations favor inward shifts, and vertical perturbations mainly suppress the polarized intensity. For the illustrative M87$^*$-like cases, the phase evolution of the $\beta_2$ polarization coefficient further traces how localized magnetic reconfiguration contributes to the global EVPA twist of the ring. These results show that horizon-scale polarization variability can provide a sensitive diagnostic of localized magnetic restructuring and that the present framework offers a physically transparent way to interpret time-dependent polarimetric signatures in future EHT observations.
\end{abstract}
\keywords{Black hole; Thin accretion disk; Polarization}

\maketitle

\section{Introduction}
\label{sec:intro}
Observations by the Event Horizon Telescope (EHT) have directly imaged horizon-scale structures around the supermassive black holes of M87$^{*}$ and Sagittarius A$^{*}$, revealing bright, asymmetric ring-like emission surrounding a central brightness depression \cite{EventHorizonTelescope:2019dse,EventHorizonTelescope:2022wkp}. Polarimetric observations further provide direct access to the magnetic field structure near the event horizon, which shows significant linear polarization fractions and coherent electric vector position angle (EVPA) patterns across the emission ring \cite{EventHorizonTelescope:2021bee,EventHorizonTelescope:2021srq,EventHorizonTelescope:2024hpu,EventHorizonTelescope:2024rju}. To interpret these observations, the EHT Collaboration has constructed extensive libraries of general relativistic magnetohydrodynamic (GRMHD) simulations and performed detailed comparisons with synthetic images. They demonstrate that the observed intensity and polarization structures can be broadly reproduced within magnetized accretion flow models \cite{EventHorizonTelescope:2019pgp,EventHorizonTelescope:2019ggy,EventHorizonTelescope:2022urf}. While these simulations provide a self-consistent description of the plasma dynamics and radiative processes, their complexity makes it difficult to isolate how specific magnetic-field configurations contribute to individual polarization features. This has motivated the development of simplified, semi-analytical models. For example, Narayan \textit{et al.} \cite{Narayan2021} introduced a polarized ring model based on synchrotron emission from a rotating plasma ring in Schwarzschild spacetime. They incorporated relativistic light bending through the Beloborodov approximation and polarization transport via the Walker--Penrose constant, then established a direct mapping between magnetic-field geometry and observable polarization signatures. This framework was subsequently extended to the Kerr spacetime \cite{Gelles2021}. It has also been widely applied in studies of modified gravity and alternative spacetime scenarios \cite{Qin:2021xvx, Delijski:2022jjj, zhang2024polarized, Zhang:2022klr, Tlemissov2025}.  Complementary approaches based on different modeling strategies and computational implementations within General Relativity have been developed to investigate polarization signatures \cite{huang2024coport,Aslam:2025hgl,Zeng:2025kyv,Wan:2025gbm}. As a recent example of these developments, Hou \textit{et al.} used a semi-analytical approach to investigate EVPA patterns in the near-horizon inner-shadow region and showed that these patterns may encode information about black hole spin and the underlying spacetime geometry~\cite{Hou:2024qqo}.

Despite the success of stationary and semi-analytical models in reproducing horizon-scale polarization features, multi-epoch observations reveal significant variability in the polarization structure. Comparison between the 2017, 2018, and 2021 EHT datasets shows substantial changes in both polarized intensity distribution and EVPA morphology \cite{EventHorizonTelescope:2025vum}. Such variability indicates that the magnetic field structure in the inner accretion flow is dynamically evolving. Time-dependent models are required to explain the variability. 
GRMHD simulations and theoretical analyses of magnetically arrested disk (MAD) models indicate that the inner accretion flow near the black hole can be highly magnetized \cite{Chael:2021rjo,Mizuno:2021esc,Yuan:2022mkw,Yang:2024kpz}. In such environments, magnetic reconnection is expected to play a critical role in driving the variability. As shown by GRMHD simulations, thin current sheets frequently form near the disk midplane and at the disk--jet interface, where magnetic shear becomes strong and intermittent reconnection events efficiently dissipate magnetic energy and accelerate particles \cite{Kagan:2014hea,Ball:2018,Nalewajko:2024qtc}. The reconnection process often produces plasmoids or magnetic flux ropes that can grow to several gravitational radii and propagate through the inner flow. GRMHD and general relativistic ray-tracing (GRRT) studies of orbiting flux tubes, flux-rope ejections, and flare-producing magnetic structures have shown that such transient features can produce time-dependent emission, light-curve variability, and, when polarized radiative transfer is included, characteristic polarization signatures near the black hole~\citep{Porth:2020txf,Najafi-Ziyazi:2023oil,Jiang:2024gtk,Jiang:2025huk}. These structures, sometimes referred to as magnetic bubbles, can perturb the local magnetic field geometry and modify the synchrotron emission pattern as they evolve \cite{Nathanail:2020wap,Ripperda:2020bpz,LinYuan:2024}. Such transient magnetic structures, governed by the regulation of magnetic memory within the inner flow \cite{Santhiya:2025ouh}, are therefore natural candidates to produce the observed variability in polarization morphology and horizon-scale flaring activity \cite{GrigorianDexter:2024}.

In this work, we address this gap by extending the polarized ring framework to incorporate time-dependent perturbations associated with reconnection-driven magnetic bubbles, including their emergence, advection, and decay. We further introduce a phenomenological magnetic-memory prescription to describe the finite-time response of the magnetic field and its relaxation toward the background configuration, and include a conservative kinetic-response closure that links the ordered magnetic-energy depletion to a local change in the emitting plasma velocity. Within this setup, we examine the resulting evolution of the Stokes $Q$ and $U$ parameters and the large-scale EVPA morphology. This approach provides a physically transparent framework for linking reconnection-driven magnetic restructuring to horizon-scale polarization variability.

This paper is organized as follows. In Sec.~II, we introduce the baseline polarized ray-tracing setup, the localized magnetic-bubble and memory prescriptions, and the associated local kinetic response. In Sec.~III, we first discuss how the reconnection-bubble perturbation evolves under different choices of the model parameters, and then investigate the resulting polarization behavior, including an illustrative M87$^*$-like application and the corresponding \(\beta_2\)-phase response. We summarize our conclusions in Sec.~IV.

\section{Models}
\label{sec:2}
 
In this section, we first summarize the unperturbed background configuration and polarized ray-tracing prescription, and then introduce the localized magnetic-bubble perturbation, the associated magnetic-memory effect, and the local kinetic-response prescription.

\subsection{Baseline polarized ray-tracing model}
\label{sec:2-1}
As an unperturbed background, we consider synchrotron emission from an equatorial emitting layer in a Schwarzschild spacetime. In units $G=c=M=1$, the metric takes the form
\begin{align}
	ds^2 = -f(r)\,dt^2 + &f(r)^{-1}dr^2 + r^2(d\theta^2+\sin^2\theta\,d\phi^2),\nonumber \\
	f(r)&=1-\frac{2}{r}.
\end{align}

Synchrotron emission is assumed to originate from plasma elements in an equatorial emitting layer. The photons emitted by the plasma propagate along null geodesics and are detected by a distant observer at inclination angle $\theta_{\rm obs}$. Following \cite{Narayan2021}, we compute the observed polarization by parallel transporting the polarization vector along the null geodesics using the conserved Walker--Penrose constant.

The emitting plasma is characterized by a steady background flow in the equatorial plane. The three-velocity of the flow is described by its magnitude and direction as \cite{Narayan2021} 
\begin{equation}
	\boldsymbol{\beta}_{0} = \beta_{0} (\cos\chi_0 \,\hat{\mathbf{x}} + \sin\chi_0 \,\hat{\mathbf{y}}),
\end{equation}
where \(\chi_0\) specifies the direction of the flow within the equatorial plane. The corresponding background Lorentz factor is \(\Gamma_0=(1-\beta_0^2)^{-1/2}\). In the unperturbed model, \(\beta_0\) and \(\Gamma_0\) are used in the local Lorentz transformations between the plasma frame and the locally static frame. In Sec.~II~B, this background value is generalized to a local Lorentz factor \(\Gamma_{\rm loc}(t,r,\phi)\) when the phenomenological kinetic response of the reconnection bubble is included.

The magnetic field in the plasma rest frame is described by a prescribed background component
\begin{equation}
	\mathbf{B}_0=(B_{r,0},\,B_{\phi,0},\,B_{z,0}),
\end{equation}
where $B_{z,0}$ denotes the component perpendicular to the equatorial plane in the local plasma frame. This field defines the reference polarization structure in the absence of time-dependent perturbations.

We specify the synchrotron emission prescription following the semi-analytical polarized ring model of Ref.~\cite{Narayan2021}. In this model, the polarized emissivity is prescribed locally in the fluid frame and then mapped to the observer's screen through the frequency-shift factor and the parallel transport of the polarization vector. Let \(\zeta\) be the pitch angle between the photon propagation direction \(\mathbf{k}_{(F)}\) and the local magnetic field \(\mathbf{B}\) in the fluid frame, defined by \cite{Narayan2021} 
\begin{equation}
	\sin\zeta =
	\frac{|\mathbf{k}_{(F)}\times \mathbf{B}|}
	{|\mathbf{k}_{(F)}|\,|\mathbf{B}|}.
\end{equation}

For a power-law spectrum $I_{\nu}\propto \nu^{-\alpha_\nu}$, the angular dependence of synchrotron emission enters through this pitch-angle factor, and the observed linearly polarized intensity is taken to scale as \cite{Narayan2021}
\begin{equation}
		|P| \propto
		\delta^{3+\alpha_\nu}
		l_p
		|\mathbf{B}|^{1+\alpha_\nu}
		\sin^{1+\alpha_\nu}\zeta ,
\end{equation}
where $\delta=1/k^{\hat t}_{(F)}$ is the frequency-shift factor. The effective path length through an emitting layer of thickness \(H\) is \cite{Narayan2021}
\begin{equation}
		l_p =
		\frac{k^{\hat t}_{(F)}}{k^{\hat z}_{(F)}}H .
\end{equation}
In our numerical calculations we set $\alpha_{\nu}=1$, as in Ref.~\cite{Narayan2021}, so that the polarized intensity reduces to
\begin{equation}
		|P| \propto
		\delta^4 l_p |\mathbf{B}|^2\sin^2\zeta .
\end{equation}
The emitting layer is assumed to be optically thin at the observing frequency, so synchrotron self-absorption is neglected. As an additional simplifying assumption, Faraday rotation and conversion are also omitted, following the polarized-ring treatment of Ref.~\cite{Narayan2021}. The local polarization direction is fixed by the usual synchrotron condition that the electric vector is orthogonal to both the photon direction and the magnetic field in the fluid frame,
\begin{equation}
		\mathbf{f}_{(F)} \propto \mathbf{k}_{(F)}\times \mathbf{B}.
\end{equation}
After parallel transport to the observer's screen, the image-plane electric-field components $E_{x,{\rm obs}}$ and $E_{y,{\rm obs}}$ are normalized such that
\begin{equation}
		E_{x,{\rm obs}}^2+E_{y,{\rm obs}}^2=|P|.
\end{equation}
The corresponding Stokes parameters are then \cite{Narayan2021}
\begin{equation}
		Q = E_{x,{\rm obs}}^2 - E_{y,{\rm obs}}^2,
		\qquad
		U = 2E_{x,{\rm obs}}E_{y,{\rm obs}},
\end{equation}
from which the complex linear polarization and EVPA are obtained as \cite{Narayan2021}
\begin{equation}
		P=Q+iU,
		\qquad
		{\rm EVPA}=\frac{1}{2}\arctan\left(\frac{U}{Q}\right).
\end{equation}

In the absence of additional perturbations, this framework produces a stationary polarization pattern determined by the background magnetic field configuration, relativistic motion of the emitting plasma, and gravitational lensing near the black hole. This steady configuration serves as the baseline state in our model. 

\subsection{Time-dependent magnetic bubble perturbations}
\label{sec:2-2}

Building on the stationary background described in Sec.~\ref{sec:2-1}, we introduce localized magnetic disturbances associated with magnetic reconnection in the accretion flow. Magnetic reconnection can fragment current sheets and generate coherent structures such as magnetic islands, plasmoids, or flux ropes. In the present work, we focus on the effective imprint of these reconnection-generated structures on polarized synchrotron emission, while the associated bulk-flow response and other energy channels are discussed below.

\begin{figure}[tbp]
	\centering
	\includegraphics[width=8cm,height=5cm]{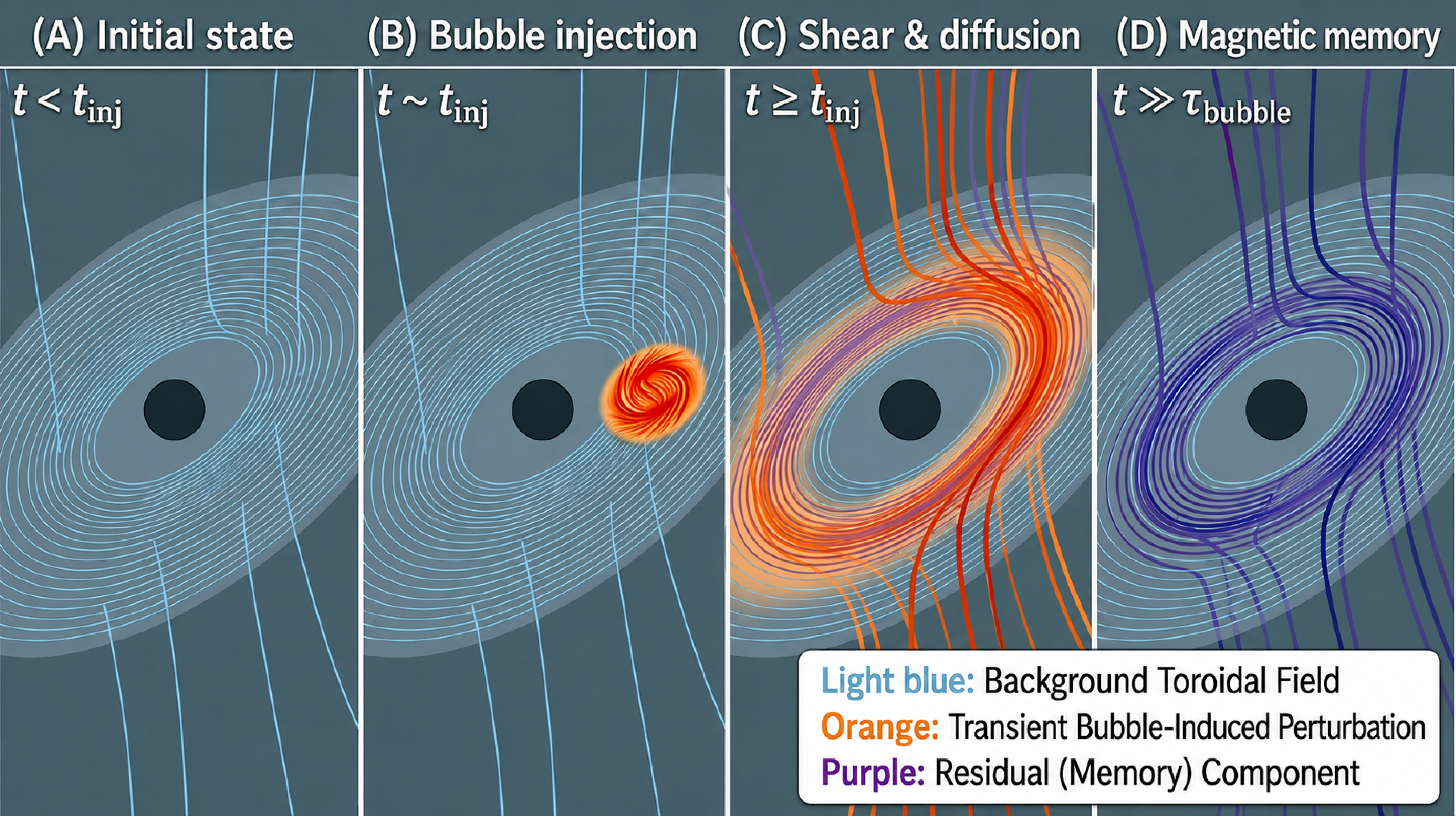}
	\caption{Schematic illustration of the reconnection-bubble model. A localized magnetic bubble is injected into the equatorial emitting region, producing magnetic depletion and local field reorientation. The perturbation advects with the orbital flow, broadens spatially, and decays on the timescale $\tau_{\rm bubble}$ while leaving an imprint on the surrounding magnetic geometry. The resulting evolution generates time-dependent polarization variability on the image plane.\label{fig:1}}
\end{figure}

In the present model, we represent these reconnection-generated structures phenomenologically as transient, spatially localized magnetic ``bubbles'' embedded in an otherwise quasi-stationary accretion flow. As illustrated schematically in Fig.~\ref{fig:1}, a magnetic bubble is introduced as a localized perturbation superposed on the stationary background field, where the field strength is temporarily weakened and the field orientation is locally distorted. After formation, the bubble is assumed to evolve under the combined influence of differential rotation, turbulent mixing, and shear in the flow, so that it advects, broadens, and gradually decays with time while leaving a finite imprint on the surrounding magnetic configuration.

We model a magnetic bubble as a localized perturbation injected at time $t_{\rm inj}$ and centered at $(r_0,\theta_0,\phi_0)$. These parameters are effective quantities that capture the characteristic spatial and temporal scales of reconnection-driven structures in the inner accretion flow, rather than serving as purely phenomenological fitting parameters. In particular, the injection radius $r_0$ is chosen to lie in the inner reconnection-active region of the flow, typically from a few $r_g$ out to $\sim 10$--$15\,r_g$, where current sheets, flux ropes, and flare-producing magnetic structures are found to emerge in simulations \citep{LinYuan:2024,GrigorianDexter:2024}. In this sense, our Gaussian envelope provides a mesoscopic description of the observable reconnection-perturbed zone projected onto the synchrotron-emitting ring, 
\begin{align}
	\mathcal{E}(t,r,\theta,\phi) &= H(t-t_{\rm inj})\,
	\exp\!\left[-\frac{t-t_{\rm inj}}{\tau_{\rm bubble}}\right]
	\exp\!\left[-\frac{(r-r_0)^2}{2\sigma_r^2(t)}\right] \nonumber\\
	&\times
	\exp\!\left[-\frac{(\theta-\theta_0)^2}{2\sigma_\theta^2(t)}\right]
	\exp\!\left[-\frac{\Delta\phi^2(t)}{2\sigma_\phi^2(t)}\right],
\end{align}
where $H(t-t_{\rm inj})$ is the Heaviside step function, which enforces that the perturbation is activated only after the injection time $t_{\rm inj}$. The parameter $\tau_{\rm bubble}$ denotes the effective lifetime of the reconnection-driven disturbance. Physically, it represents the time interval over which the localized magnetic depletion remains dynamically relevant before being erased by advection, shear, and dissipation. GRMHD-based flare models suggest that the evolution times of flux ropes and reconnection-powered structures in the inner flow can range from local transport times in order $\sim 10\,r_g/c$ to $\sim 10^2\,r_g/c$, depending on the magnetic geometry and dynamical state of the system \citep{LinYuan:2024,GrigorianDexter:2024}. It is therefore natural to regard $\tau_{\rm bubble}$ as an effective parameter of $0\sim 1$ 
of the local orbital time at $r_0$. The azimuthal offset $\Delta\phi(t)$ measures the separation relative to the time-dependent bubble center, which is assumed to advect with the local Keplerian angular frequency. Explicitly, we write
\begin{equation}
	\Delta\phi(t) = \phi - \left[\phi_0 + \Omega_K(r_0)(t - t_{\mathrm{inj}})\right],
\end{equation}
where $\phi_0$ is the initial azimuthal location of the bubble center and $\Omega_K(r)=r^{-3/2}$ is the Keplerian angular frequency in Schwarzschild spacetime. This prescription reflects the expectation that a reconnection-generated magnetic structure is advected azimuthally by the background flow, following approximately circular motion while remaining spatially localized over a finite dynamical timescale.

The spatial extent of the bubble evolves with time due to turbulent mixing, shear, and orbital stretching in the surrounding accretion flow. We model this using
\begin{equation}
	\sigma_i^2(t)=\sigma_{i,0}^2+2D_i(t-t_{\rm inj}), \qquad i\in\{r,\theta,\phi\},
\end{equation}
where $\sigma_{i,0}$ are the initial widths and $D_i$ are effective broadening parameters. This prescription captures the growth of the reconnection-perturbed region under turbulence and shear, consistent with GRMHD simulations \citep{LinYuan:2024,GrigorianDexter:2024}. Notably, while the current sheet itself remains locally thin, the associated plasmoid- or flux-rope-driven disturbance can extend to a substantially larger region \citep{Ball:2018,LinYuan:2024,GrigorianDexter:2024}.

To encode the magnetic depletion produced by reconnection, we introduce a residual amplitude parameter $A_{0}<1$, which measures the residual magnetic-field amplitude that remains after the local release of magnetic energy, and set the post-reconnection field strength as
\begin{equation}
	|B_{\rm bubble}| = A_{0}\,|B_{0}|.
\end{equation}

Here \(A_0\) is a phenomenological depletion parameter that controls the residual amplitude of the ordered magnetic field after reconnection, rather than an energy-partition coefficient. In this sense, the local magnetic-energy release is represented in the polarized-emission calculation through the reduced post-reconnection field amplitude \(A_0|\mathbf B_0|\). Since the magnetic energy density of the ordered field scales with \(|\mathbf B|^2\), a reduction of the ordered-field amplitude represents that part of the local magnetic energy has been converted out of the ordered magnetic component. The model therefore tracks the magnetic-side imprint of reconnection, namely the weakening and reorganization of the ordered field that directly enters the synchrotron polarization calculation. This choice is physically motivated by GRMHD studies showing that reconnection-active regions can contain substantial magnetic energy and undergo significant local weakening of magnetic support during dissipation~\citep{Ball:2018,LinYuan:2024}.

Beyond amplitude reduction, magnetic reconnection generically induces a redistribution of magnetic-field directions that is not instantaneous, but instead relaxes over a finite interval as newly reconnected field lines retract, shear, and interact with the surrounding turbulent flow. To capture this behavior, we introduce magnetic-memory variables that encode the time-integrated response of the system to localized perturbations. We define a corresponding memory variable $M_i(t,r,\phi)$ on the equatorial emission layer, where  $i\in\{r,\phi,z,\chi\}$.  $i=r,\phi,z$ denote the components of the magnetic field defined with respect to a local orthonormal basis and $i=\chi$ denotes the fluid-velocity orientation angle.  $M_i$ is constructed as a causal convolution of a local source term $S_i$ with an exponential memory kernel,
\begin{equation}
	M_i(t,r,\phi)=f_{{\rm Mem},i}\int_{0}^{t}S_i(t',r,\phi)\,
	\exp\!\left[-\frac{t-t'}{\tau_{\rm Mem}}\right]dt',
\end{equation}
where $\tau_{\rm Mem}$ denotes an effective magnetic-memory timescale, i.e., the interval over which the flow retains the imprint of a reconnection event before turbulent mixing and shear erase the perturbation.  It represents a mesoscopic relaxation timescale for post-reconnection magnetic reorganization. It is longer than the local sheet-crossing time but comparable to or shorter than the characteristic orbital or shear time in the inner flow. Thus $\tau_{\rm Mem}$ characterizes the finite-time evolution of the large-scale magnetic geometry rather than any microscopic plasma timescale. The coefficients $f_{{\rm Mem},i}$ regulate how efficiently a reconnection event imprints residual reorientation onto different degrees of freedom, thereby encoding the directional bias of the post-reconnection magnetic configuration. This interpretation is physically motivated by GRMHD studies of current sheets, which show that the outcome of reconnection depends sensitively on the magnetization, plasma-$\beta$, and guide-field strength, with MAD-like systems often exhibiting guide fields comparable to the reconnecting field itself \citep{Ball:2018}. In this sense, different choices of $f_{{\rm Mem},r}$, $f_{{\rm Mem},\phi}$, and $f_{{\rm Mem},z}$ correspond to different effective reconnection geometries rather than merely formal component weights.

The source terms $S_i$ are constructed as the product of the bubble envelope evaluated on the equatorial emission surface and a time-dependent oscillatory factor, i.e.
\begin{equation}
	S_i(t,r,\phi)=\mathcal{E}(t,r,\pi/2,\phi)\,
	O_i\!\left[\omega_i (t-t_{\rm inj})+\psi_i\right],
\end{equation}
where $O_i$ denotes either a sine or cosine function. The restriction to $\theta=\pi/2$ follows the semi-analytical framework of \cite{Narayan2021}, in which the observed radiation is modeled as originating from an effective, optically thin emitting ring in the equatorial plane. In this description, the emission layer serves as a reduced representation of a finite-thickness radiating region, so that external perturbations enter the observable signal through their effective imprint on this surface. The oscillatory term provides a minimal representation of the time-dependent reconfiguration of the magnetic field associated with reconnection, including repeated formation and relaxation of magnetic structures. The frequencies $\omega_i$ are taken to be proportional to $\Omega_K(r_0)$ with dimensionless coefficients of order unity that vary between components, thereby introducing multiple variability timescales associated with orbital motion, shear, and magnetic stresses. The phase offsets $\psi_i$ set the initial state of each component at $t_{\rm inj}$ and ensure that different components evolve out of phase, so that the perturbation leads to a genuine reorientation of the magnetic field rather than a purely scalar modulation.

At each spacetime point on the emission surface, the memory variables define a normalized preferred magnetic direction associated with the perturbation,
\begin{equation}
	\hat{\mathbf{b}}_{\rm bubble}
	=
	\frac{(M_r,M_\phi,M_z)}
	{\sqrt{M_r^2+M_\phi^2+M_z^2}},
\end{equation}
where the relative weights of $(M_r,M_\phi,M_z)$ determine the orientation of the post-reconnection magnetic configuration, while the overall perturbation strength is controlled separately through the prescribed magnetic depletion factor.

The local magnetic field on the emission surface is then obtained by smoothly blending the background field with the perturbation-induced field,
\begin{equation}
	\mathbf{B}(t,r,\phi)=(1-w)\mathbf{B}_0+wA_0|\mathbf{B}_0|\,\hat{\mathbf{b}}_{\rm bubble},
\end{equation}
where $w=\mathcal{E}(t,r,\pi/2,\phi)$ ensures a continuous transition between unperturbed regions and perturbation-dominated regions. Equivalently, writing \(M=\sqrt{M_r^2+M_\phi^2+M_z^2}\), the individual magnetic-field components entering the polarized emission calculation are
	\begin{equation}
		B_i(t,r,\phi)
		=
		(1-w)B_{i,0}
		+
		wA_0|\mathbf{B}_0|\frac{M_i}{M},
		\qquad i=r,\phi,z .
	\end{equation}
	
This prescription does not add an independent magnetic component on top of the background field. Instead, it constructs the local field as a weighted transition between the background configuration and a post-reconnection field whose direction is determined by the memory vector \((M_r,M_\phi,M_z)\). The envelope \(w\) controls the local influence of the bubble, \(A_0|\mathbf B_0|\) sets the residual ordered-field amplitude, and the ratios \(M_r/M\), \(M_\phi/M\), and \(M_z/M\) determine the relative radial, azimuthal, and vertical contributions. For example, if the background field is initially dominated by its radial component, a decrease in \(M_r/M\) together with an increase in \(M_z/M\) rotates the preferred post-reconnection direction out of the equatorial plane and toward the vertical direction. In the limit \(M_z\gg M_r,M_\phi\), the reorganized field becomes locally poloidal. More generally, different relative memory components describe different effective post-reconnection magnetic geometries on the emitting layer.

Magnetic reconnection can redistribute the released magnetic energy among coherent bulk motion, plasma heating, nonthermal particle acceleration, and turbulence~\cite{Sironi:2025kgn}. The present work is designed to isolate the polarization signatures produced by the reorganization and depletion of the ordered magnetic field, together with the associated change in the bulk motion of the emitting plasma. Within the semi-analytical polarized-ring framework adopted here, the synchrotron emissivity is prescribed directly from the local magnetic field, pitch angle, and relativistic frequency shift, rather than being derived from an independently evolved electron temperature or particle distribution. A self-consistent treatment of reconnection heating and nonthermal acceleration would therefore require additional evolution equations for the plasma internal energy and particle distribution, as well as a corresponding thermal or nonthermal synchrotron-emission prescription. Turbulent dissipation would likewise require an explicit model for the unresolved magnetic-field fluctuations. These additional ingredients would introduce physical effects beyond the geometric and kinematic polarization response considered in the present work. We therefore do not follow the complete energy partition, but retain the coherent bulk-flow channel that can be incorporated consistently into the existing polarized ray-tracing framework.

We allow the local speed magnitude to respond to the depletion of the ordered magnetic field through a simplified proportional prescription. This retains a direct connection between the released ordered magnetic energy and the kinematic response of the emitting plasma without introducing additional thermodynamic or particle-evolution equations. Using the same bubble weight \(w\) defined above, we define the local residual field-amplitude factor as
\begin{equation}
		A_{\rm loc}(t,r,\phi)
		=
		1-\left(1-A_0\right)w(t,r,\phi).
\end{equation}

The quantity \(A_{\rm loc}\) describes the local residual fraction of the ordered magnetic-field amplitude. Since the ordered magnetic-energy density scales quadratically with the field amplitude, the corresponding fractional depletion of ordered magnetic energy is
\begin{equation}
		\mathcal R_B(t,r,\phi)
		=
		\max\left[
		0,\,
		1-A_{\rm loc}^2(t,r,\phi)
		\right].
\end{equation}

The quantity \(\mathcal R_B\) serves as an amplitude-based magnetic-energy depletion tracer associated with the prescribed reduction of the ordered field amplitude. In this prescription, the amplitude depletion and the directional reorganization are encoded by two complementary ingredients: \(\mathcal R_B\) controls the kinetic response to the released ordered magnetic component, while the memory vector \((M_r,M_\phi,M_z)\) controls the post-reconnection field direction. We assign a fraction \(\epsilon_{\rm kin}\) of the released ordered magnetic-energy component to coherent bulk kinetic motion. Kinetic studies of relativistic reconnection show that, although the fastest outflows can approach the Alfv\'enic limit, most of the plasma flows out at trans-relativistic speeds and most of the dissipated magnetic energy is converted into internal plasma motions and heating rather than coherent bulk acceleration~\cite{Sironi:2025kgn}. We therefore treat the coherent bulk-kinetic channel as subdominant and adopt the conservative phenomenological value $\epsilon_{\rm kin}=0.1.$

The second parameter, \(\xi_B\), controls how strongly a given fractional depletion of ordered magnetic energy changes the bulk Lorentz factor. Physically, it represents the effective amount of ordered magnetic energy available per unit plasma inertia. In the present semi-analytical model, the physical density, temperature, and enthalpy of the emitting plasma are not evolved independently. Therefore, \(\xi_B\) is introduced as an effective normalization that absorbs the physical magnetic-field strength and plasma inertia that are not specified by the polarized-ring model.

This interpretation can be connected to the enthalpy-normalized magnetization commonly used in relativistic reconnection theory. If \(B_{\rm phys,0}\) denotes the physical reconnecting magnetic-field strength and \(h\) denotes the relativistic plasma enthalpy density, including rest-mass, internal-energy, and pressure contributions, one defines
\begin{equation}
		\sigma_h
		=
		\frac{B_{\rm phys,0}^2}{4\pi h}.
\end{equation}

The corresponding ordered magnetic-energy density is \(u_{B,0}^{\rm ord}=B_{\rm phys,0}^2/(8\pi)\). Thus the ordered magnetic energy per unit enthalpy is \(u_{B,0}^{\rm ord}/h=\sigma_h/2\). In this sense, \(\xi_B\) can be regarded as the effective magnetic-energy-to-inertia ratio encoded in the toy model,
\begin{equation}
		\xi_B
		\sim
		\frac{u_{B,0}^{\rm ord}}{h}
		\simeq
		\frac{\sigma_h}{2}.
\end{equation}

Thus larger \(\xi_B\) corresponds to a more strongly magnetized or lower-inertia reconnecting region, for which the same fractional magnetic-energy depletion produces a larger bulk acceleration. In the calculations below, we adopt $\xi_B=0.5$, which corresponds within this prescription to \(u_{B,0}^{\rm ord}/h\simeq0.5\), or equivalently to an enthalpy-normalized magnetization of order unity, \(\sigma_h\simeq1\). We use this value as a representative fiducial normalization rather than as a quantity calibrated to a particular GRMHD simulation. The Lorentz-factor increment associated with the phenomenological kinetic response is then prescribed as
\begin{equation}
		\Delta\Gamma(t,r,\phi)
		=
		\epsilon_{\rm kin}\xi_B\mathcal R_B(t,r,\phi).
\end{equation}

This expression shows explicitly that the local acceleration is controlled by three factors: the fractional ordered magnetic-energy depletion \(\mathcal R_B\), the fraction \(\epsilon_{\rm kin}\) of the released ordered magnetic energy assigned to coherent bulk motion, and the effective magnetic-energy-to-inertia normalization \(\xi_B\). The local Lorentz factor and speed are then
\begin{align}
		\Gamma_{\rm loc}(t,r,\phi)
	&
		=
		\Gamma_0
		+
		\epsilon_{\rm kin}\xi_B\mathcal R_B(t,r,\phi),
	\\
		\beta_{\rm loc}(t,r,\phi)
	&
		=
		\sqrt{1-\Gamma_{\rm loc}^{-2}(t,r,\phi)}.
\end{align}

The fixed-speed limit is recovered by setting \(\epsilon_{\rm kin}=0\). The reconnection event can also perturb the local direction of the fluid velocity through the velocity-memory variable
\begin{equation}
		\chi(t,r,\phi)
		=
		\chi_0+M_\chi(t,r,\phi),
\end{equation}
where \(M_\chi\) denotes the memory-driven correction to the background flow angle. The full local velocity entering the Lorentz transformation is therefore
\begin{equation}
		\boldsymbol{\beta}_{\rm loc}(t,r,\phi)
		=
		\beta_{\rm loc}(t,r,\phi)
		\left[
		\cos\chi(t,r,\phi)\,\hat{\mathbf x}
		+
		\sin\chi(t,r,\phi)\,\hat{\mathbf y}
		\right].
\end{equation}

In the polarized ray-tracing calculation, \(\beta_{\rm loc}\) and \(\Gamma_{\rm loc}\) replace the background constants \(\beta_0\) and \(\Gamma_0\) in the local Lorentz transformations. The kinetic response modifies the locally emitted polarization through Doppler boosting, aberration, the pitch-angle factor \(\sin\zeta\), and the local polarization vector, which is then parallel transported to the observer's screen.

Taken together, this framework provides an internally consistent effective description of reconnection-driven magnetic perturbations, incorporating magnetic weakening, directional redistribution, finite memory, spatial spreading, and a conservative local kinetic response on the emission surface. It is therefore well suited for exploring time-dependent polarized emission from black-hole accretion flows, where localized reconnection activity can imprint observable polarization signatures that persist beyond the initial energy-release episode.

\subsection{Parameter Exploration}
\label{sec:2-3}
Before presenting the polarization results, we clarify the physical status of the key parameters appearing in the reconnection-bubble model. To illustrate the roles of the depletion amplitude $A_0$ and the decay (recovery) timescale $\tau_{\rm bubble}$ in controlling the temporal evolution of the magnetic field, we examine the magnetic-field strength at the bubble center normalized to the background value. In our prescription, this quantity can be approximated as
\begin{equation}
	\frac{B(t)}{B_0} \simeq (1-w_{t}) + w_{t}A_0 ,
\end{equation}
where the temporal weight function is given by
\begin{equation}
	w_{t}(t)=\exp[-(t-t_{\rm inj})/\tau_{\rm bubble}], \qquad t\ge t_{\rm inj}.
\end{equation}

This simplified expression isolates the amplitude recovery at the bubble center and is intended only to illustrate the roles of $A_0$ and $\tau_{\rm bubble}$, independent of directional reorientation effects. At the injection time $t=t_{\rm inj}$, one has $w_{t}=1$, so that $B/B_0\simeq A_0$. As time proceeds, $w_{t}$ decays exponentially toward zero and the field gradually relaxes back toward the background level. 

Figure~2 illustrates how these two parameters control the recovery of the magnetic field at the bubble center. In panel~(a), the injection time is fixed at $t_{\rm inj}=1$ and the decay timescale at $\tau_{\rm bubble}=12$, while $A_{0}$ varies from $0.2$ to $1.0$. Smaller values of $A_{0}$ imply a stronger local magnetic depletion and therefore produce a deeper initial suppression of $B(t)/B_{0}$. As the perturbation decays, the field strength gradually relaxes back toward the background level. Panel~(b) highlights the effect of the decay timescale. Here the depletion amplitude is fixed at $A_{0}=0.4$, while $\tau_{\rm bubble}$ takes the values $6$, $12$, $18$, $24$, and $30$. Larger values of $\tau_{\rm bubble}$ lead to slower field recovery, hence longer-lived perturbations. Physically, this parameter characterizes the dynamical persistence of the reconnection-driven disturbance in the emitting region, consistent with the expectation from GRMHD studies that reconnection-generated magnetic structures in black-hole accretion flows evolve over finite dynamical times rather than disappearing instantaneously \citep{Ball:2018,LinYuan:2024,GrigorianDexter:2024}.

In addition to its temporal decay, the reconnection bubble is allowed to undergo spatial broadening over time. As a representative example, we consider the radial width, which evolves as
\begin{equation}
	\sigma_r^2(t)=\sigma_{r,0}^2+2D_r(t-t_{\rm inj}), \qquad t\ge t_{\rm inj},
\end{equation}
where $D_r$ determines how rapidly the perturbed region expands in the radial direction. Since the polar and azimuthal widths follow a similar diffusion-like law, the radial case is sufficient to illustrate the general behavior. Figure~3 shows the resulting evolution of $\sigma_r(t)$ for several values of $D_r$. Larger values of $D_r$ lead to a faster growth of the bubble width and hence to a more rapid spatial redistribution of the reconnection signature. Physically, this trend reflects the expectation that reconnection-driven magnetic structures do not remain confined to their initial location, but progressively spread through the surrounding flow under the combined action of mixing, shear, and orbital distortion, consistent with GRMHD-based studies of the inner accretion flow \citep{LinYuan:2024,GrigorianDexter:2024}.

To further illustrate the dynamical evolution of the reconnection bubble, we present snapshots of the bubble envelope $\mathcal{E}(t,r,\phi)$ in the equatorial plane at several representative epochs after the injection event. In these examples, the bubble center advects azimuthally with the local Keplerian angular velocity $\Omega_K(r_0)$, following the orbital motion of the background flow. At the same time, the envelope amplitude decays exponentially on the timescale $\tau_{\rm bubble}$, while the spatial widths broaden according to the adopted diffusion coefficients. Figure~\ref{fig:4} visualizes the resulting morphology in the $(r,\phi)$ plane. As time progresses, the bubble spreads over a progressively larger region of the disk while the central perturbation weakens, reflecting the gradual redistribution and dissipation of magnetic energy.

\begin{figure*}[tbp]
	\centering
	\includegraphics[width=6.5cm,height=5cm]{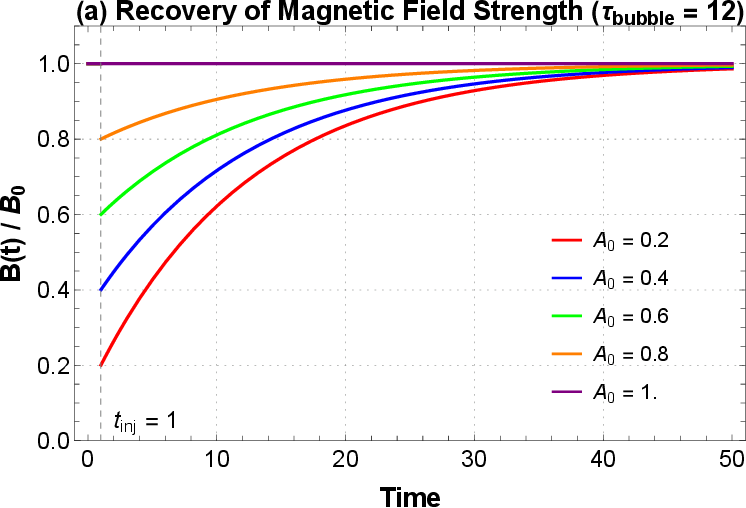}
	\includegraphics[width=6.5cm,height=5cm]{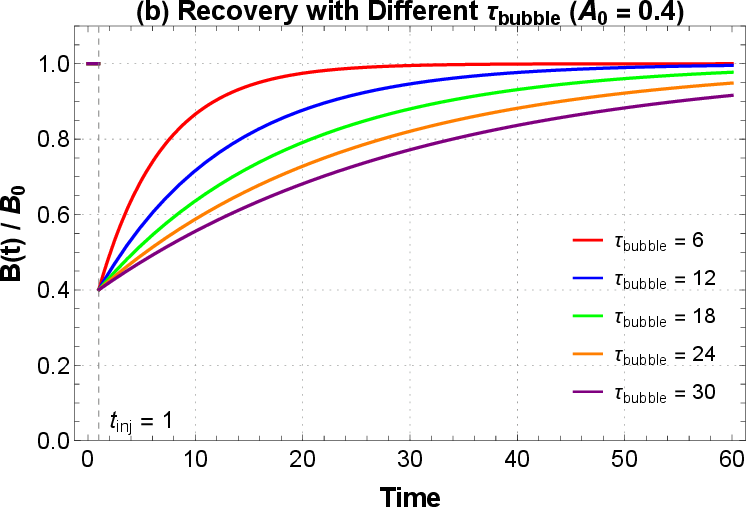}
	\caption{Evolution of the normalized magnetic-field strength at the bubble center, illustrating the ``recovery'' from a depleted reconnection region back to the background field. Panel (a) shows $B(t)/B_0$ for different depletion amplitudes $A_0$ at fixed $\tau_{\rm bubble}=12$, while panel (b) shows the effect of varying $\tau_{\rm bubble}$ at fixed $A_0=0.4$. The vertical dashed line marks the injection time $t_{\rm inj}=1$.\label{fig:2}}
\end{figure*}

\begin{figure}[tbp]
	\centering
	\includegraphics[width=6.5cm,height=5cm]{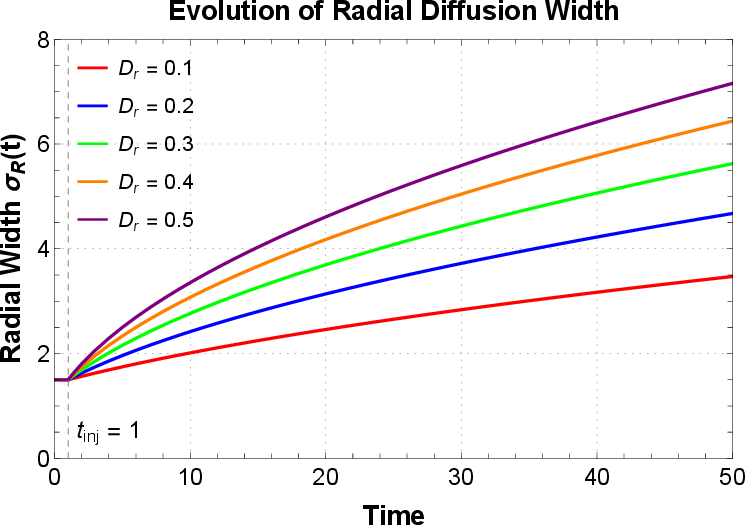}
	\caption{Time evolution of the radial width $\sigma_R(t)$ of the magnetic reconnection bubble for different effective diffusion coefficients $D_r$ (with injection time $t_{\rm inj}=1$ and initial width $\sigma_{R,0}=1.5$). The dashed vertical line marks $t_{\rm inj}$.\label{fig:3} }
\end{figure}

\begin{figure*}[tbp]
	\centering
	\includegraphics[width=5.8cm,height=4.5cm]{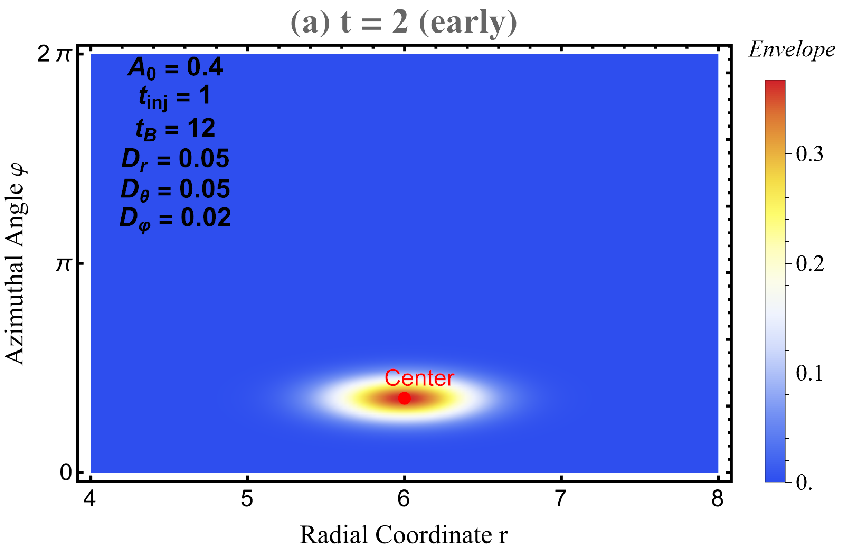}
	\includegraphics[width=5.8cm,height=4.5cm]{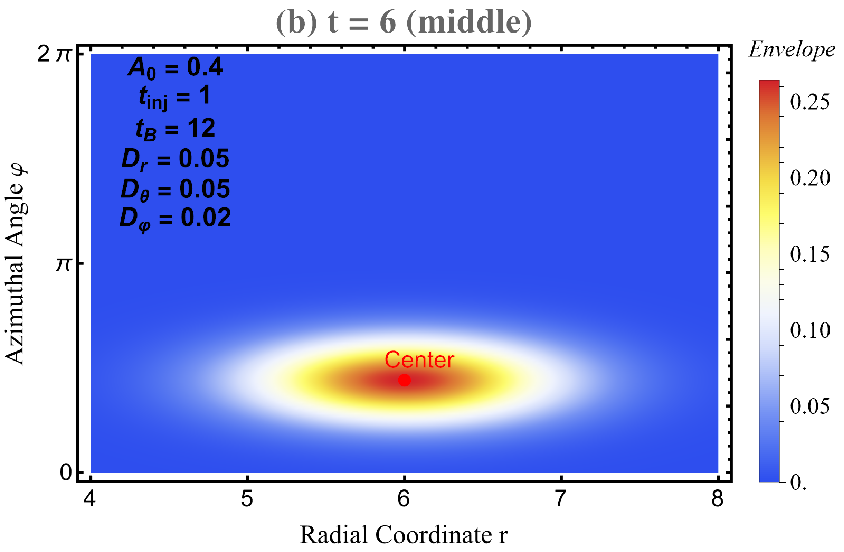}
	\includegraphics[width=5.8cm,height=4.5cm]{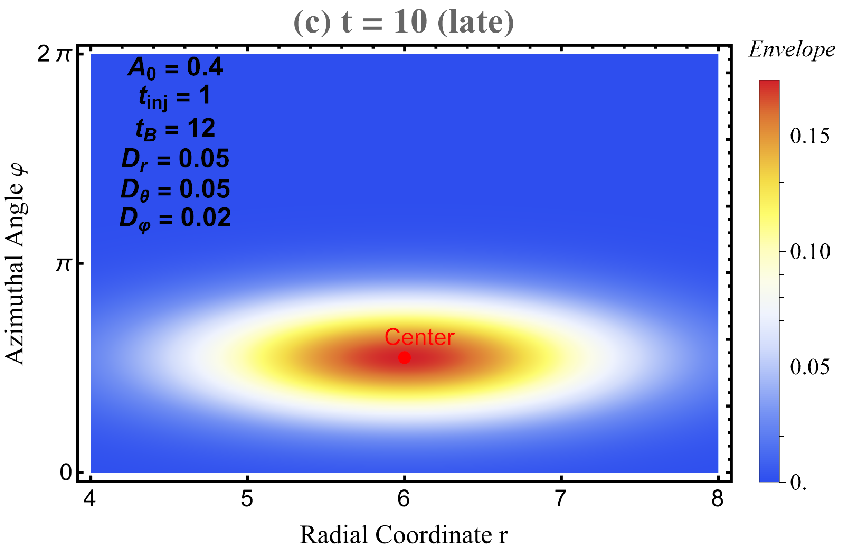}
	\caption{Snapshots of the reconnection-bubble envelope $\mathcal{E}(t,r,\phi)$ in the equatorial plane at (a) $t=2$, (b) $t=6$, and (c) $t=10$ (with $t_{\rm inj}=1$). The peak amplitude decays as $\propto A_0\exp[-(t-t_{\rm inj})/\tau_{\rm bubble}]$, while the spatial widths broaden as $\sigma_i^2(t)=\sigma_{i,0}^2+2D_i(t-t_{\rm inj})$.\label{fig:4}}
\end{figure*}

To illustrate the role of the magnetic-memory timescale in a simplified setting, we consider a representative scalar memory variable $M(t)$ governed by
\begin{equation}
	\frac{dM}{dt}=f_{\rm Mem}S(t)-\frac{M}{\tau_{\rm Mem}}.
\end{equation}

Figure~5 shows the evolution of $M(t)$ for several representative values of $\tau_{\rm Mem}$. Smaller $\tau_{\rm Mem}$ leads to a response that rises and decays more rapidly, whereas the larger $\tau_{\rm Mem}$ produces a broader and more persistent response. As seen in Fig.~5, increasing $\tau_{\rm Mem}$ mainly broadens the temporal profile and extends its trailing decay, rather than simply rescaling the overall amplitude. The figure therefore shows that $\tau_{\rm Mem}$ does not primarily control the instantaneous strength of the perturbation, but instead governs how long the post-reconnection reorganization remains dynamically relevant. This delayed relaxation is important for the polarization evolution, because it allows the magnetic geometry to retain a finite-time imprint even after the bubble amplitude itself has already started to decay.

To connect the effective bubble description with a quantity more directly relevant for polarized emission, we examine the magnitude of the total magnetic field, $|\mathbf{B}(t,r,\phi)|$, in the equatorial plane. Figure~6 shows its spatiotemporal evolution at the representative radius $r=11$. Following the injection at $t_{\rm inj}$, the field strength develops a localized depression near the bubble center, marking the onset of reconnection-driven magnetic depletion. As the disturbance evolves, this depleted region broadens and drifts azimuthally under the combined effects of bubble spreading, advection, and magnetic memory. At the same time, the perturbation near the original injection site gradually weakens, and the local field begins to recover toward the background level as the bubble decays. The resulting evolution therefore reflects two simultaneous processes: the redistribution of the reconnection signature over a wider portion of the emitting flow, and the gradual relaxation of the disturbance at the initial reconnection site. At later times, the magnetic pattern is characterized by a weaker but more extended residual perturbation, indicating that the local energy-release episode has already begun to relax even though its imprint on the surrounding field geometry is still being redistributed across the flow. This behavior is consistent with the general picture emerging from GRMHD studies, in which reconnection-generated flux ropes and magnetized bubbles continue to evolve after formation, while the field in the original reconnection region does not remain permanently depleted \citep{LinYuan:2024,GrigorianDexter:2024}. Figure~6 therefore provides a useful intermediate diagnostic between the effective bubble model and the polarization calculations presented below, showing how a localized reconnection event can generate a time-dependent magnetic pattern whose spatial imprint persists even as the original depletion begins to relax.

\begin{figure}[tbp]
	\centering
	\includegraphics[width=6.5cm,height=5cm]{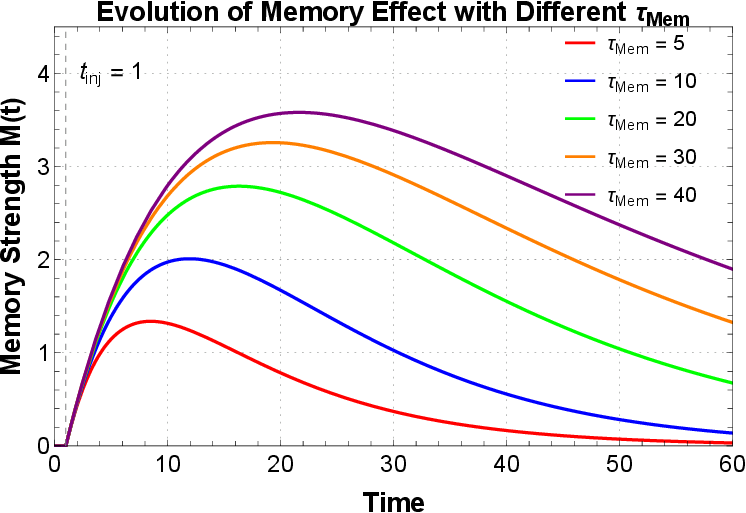}
	\caption{Evolution of the magnetic-memory variable $M(t)$ obtained from $dM/dt=f_{\rm Mem}S(t)-M/\tau_{\rm Mem}$ with $S(t)\propto \exp[-(t-t_{\rm inj})/\tau_{\rm bubble}]$ for $t\ge t_{\rm inj}$. Curves correspond to different memory timescales $\tau_{\rm Mem}$ (with $t_{\rm inj}=1$, $\tau_{\rm bubble}=12$, and $f_{\rm Mem}=0.5$).\label{fig:5} }
\end{figure}

\begin{figure}[tbp]
	\centering
	\includegraphics[width=8cm,height=4.5cm]{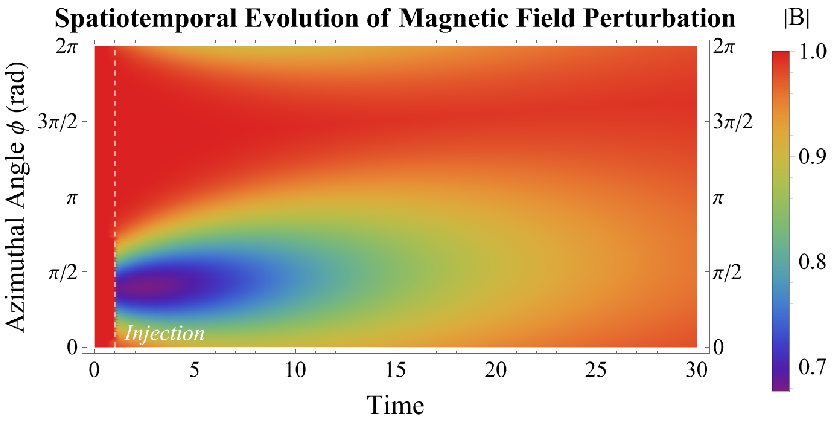}
	\caption{Spatiotemporal evolution of the magnetic-field magnitude $|\mathbf{B}(t,r,\phi)|$ in the equatorial plane at fixed radius $r=11$ as a function of time and azimuthal angle $\phi$. The field is obtained from the mixing prescription $\mathbf{B}=(1-w)\mathbf{B}_0+w\mathbf{B}_{\rm bubble}$, with $\mathbf{B}_{\rm bubble}$ determined by the memory variables. The dashed line marks the injection time $t_{\rm inj}=1$.\label{fig:6} }
\end{figure}

\section{Results: polarization images and quantitative diagnostics}
\label{sec:3}

\subsection{Component-dependent polarization response of the reconnection bubble}
\label{sec:3-1}
Using the polarized ray-tracing framework together with the time-dependent reconnection-bubble model, we now examine how localized magnetic reconfiguration imprints itself on observable polarization signatures. We focus  on how the image morphology, azimuthal EVPA structure, and global polarization response change when the post-reconnection perturbation is dominated by different magnetic-field components. By holding the background spacetime, viewing geometry, and large-scale flow properties fixed, the comparison isolates the polarization imprint of reconnection geometry itself.

Figure~7 presents three representative realizations in which the post-reconnection response is dominated by the radial (top row), azimuthal (middle row), and vertical (bottom row) magnetic-field components, respectively. Apart from the dominant memory component, all other parameters are kept fixed across the three cases. In the fiducial setup shown here, the bubble is injected at $t_{\rm inj}=1$ near $r_{\rm inj}=11$, with $\tau_{\rm bubble}=12$ and $\tau_{\rm Mem}=20$. These values should be regarded as representative inner-flow choices rather than uniquely calibrated GRMHD outputs. The two timescales are chosen to be of order the local dynamical time, with $\tau_{\rm bubble}<\tau_{\rm Mem}$ so that the magnetic depletion begins to decay before the post-reconnection field reorganization is fully erased. The memory amplitude is set to $f_{\rm Mem}=1$ as an order-unity normalization, while the velocity--memory coupling is fixed at $f_{{\rm Mem},\chi}=0.3$ as a conservative subdominant choice, allowing the flow-direction response to remain present without overwhelming the magnetic reorientation itself. Each row contains four diagnostic panels. The first column shows representative polarized images at three epochs ($t=0$, $5$, and $20$), illustrating how the polarization morphology evolves as the reconnection bubble propagates and relaxes. The second and third columns show azimuthal profiles measured along a ring at $R=11$. The second column displays the intensity variation, $\Delta I \equiv I_t-I_{t=0}$, which traces how the bubble-induced magnetic disturbance redistributes the synchrotron emissivity. The third column shows the change in polarization angle, $\Delta{\rm EVPA} \equiv {\rm EVPA}_t-{\rm EVPA}_{t=0}$, which is particularly sensitive to magnetic-field reorientation and to the relative contribution of different field components projected along the line of sight. The fourth column shows the corresponding trajectory in the \((Q,U)\) plane, providing a compact representation of the global polarization response as the system departs from the background state. Although the EVPA maps in Fig.~7 retain a coherent large-scale morphology, this does not indicate that the polarization is insensitive to the magnetic-field configuration. The comparison differs from that of Ref.~\cite{Narayan2021}, where different magnetic geometries are imposed over the emitting ring as a whole. Here the background magnetic field, viewing geometry, spacetime, and large-scale flow are kept fixed, while only a limited azimuthal sector is modified by the localized reconnection bubble. Most of the emitting ring therefore remains controlled by the same coherent background field and retains the large-scale EVPA morphology of the unperturbed state. The adopted low inclination further preserves this regular projected structure. Moreover, the bubble envelope and memory variables vary smoothly across the affected region, so the local magnetic field is reorganized coherently rather than randomly. The perturbation consequently appears as a localized EVPA displacement superposed on the common background pattern. The component dependence is therefore more clearly revealed by the residual quantities \(\Delta I\) and \(\Delta{\rm EVPA}\), together with the Stokes-\((Q,U)\) loci. Once the local kinetic response is included, \(\Delta I\) reflects the combined effects of ordered-field depletion, field reorientation, Doppler boosting, aberration, and the pitch-angle factor. In the examples shown here, all three configurations exhibit local polarized-intensity suppression, whereas the sign and amplitude of the EVPA residual remain strongly dependent on the dominant magnetic-memory component.

When the radial component dominates, the main signature is a moderate local decrease in polarized intensity together with a negative \(\Delta{\rm EVPA}\). The post-reconnection memory direction is mainly radial, so the projected magnetic-field geometry in the disturbed sector rotates in a way that shifts the local EVPA inward relative to the background state. The ordered-field depletion directly weakens the magnetic-amplitude contribution to the polarized emissivity, while the associated kinetic response further modifies the Doppler factor, aberration, and pitch-angle factor. The resulting Stokes-\((Q,U)\) trajectory is displaced from the unperturbed locus, indicating a coherent change in the integrated polarization state.

When the azimuthal component dominates, the EVPA response changes sign. The azimuthally weighted memory direction leaves a stronger toroidal imprint in the projected post-reconnection field geometry, producing a positive \(\Delta{\rm EVPA}\) over the affected azimuthal range. The polarized intensity is nevertheless still locally reduced, although more weakly than in the radial and vertical cases, because the ordered-field depletion and the kinematic correction both enter the polarized-emission calculation. Thus the azimuthal case is characterized by a positive EVPA residual together with mild polarized-intensity suppression. Its Stokes-\((Q,U)\) trajectory remains distinct from the radial case, reflecting the opposite EVPA response.

The vertical-dominated case is mainly characterized by the strongest polarized-intensity suppression among the three examples. Its EVPA residual is comparatively small and negative, showing that a vertical memory direction changes the projected polarization angle less efficiently than the radial or azimuthal cases in this viewing configuration. The strong decrease in polarized intensity arises because the vertical perturbation reduces the effective transverse magnetic-field component contributing to synchrotron polarization and changes the pitch-angle factor. Together with the local kinetic response, this produces a pronounced negative \(\Delta I\) signal while preserving a relatively coherent large-scale EVPA morphology. These results show that the three memory configurations can be distinguished by the combined behavior of polarized-intensity suppression, the sign and amplitude of \(\Delta{\rm EVPA}\), and the Stokes-\((Q,U)\) evolution.

\begin{figure*}[tbp]
	\centering
	\includegraphics[width=4cm,height=4cm]{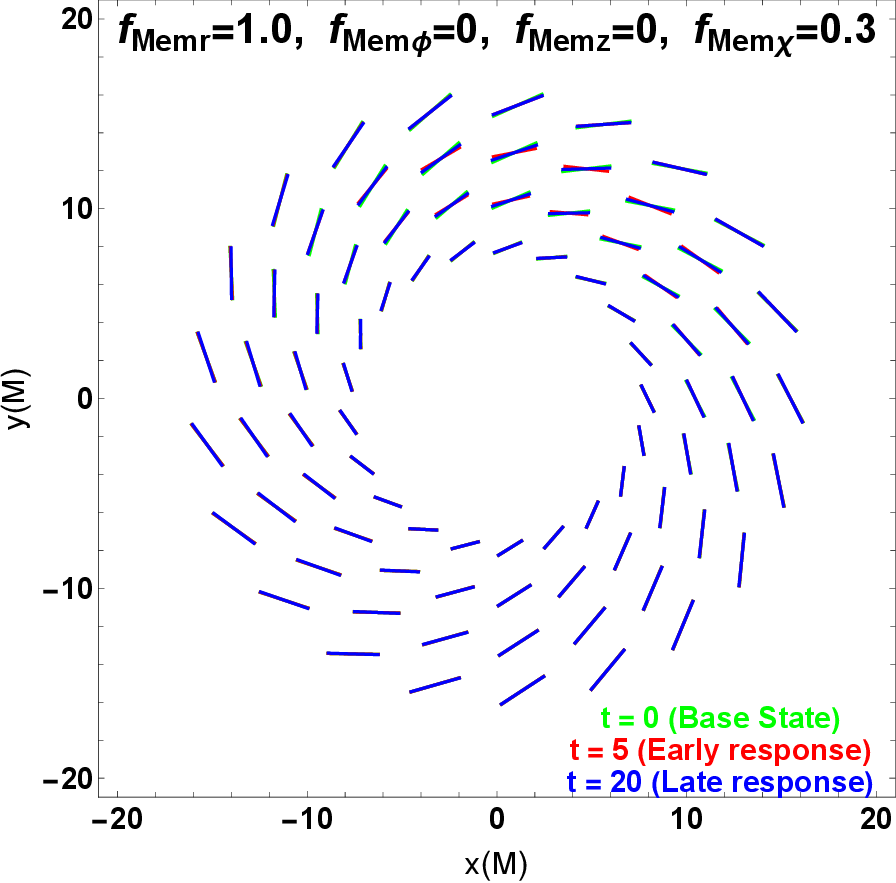}
	\includegraphics[width=4cm,height=3.5cm]{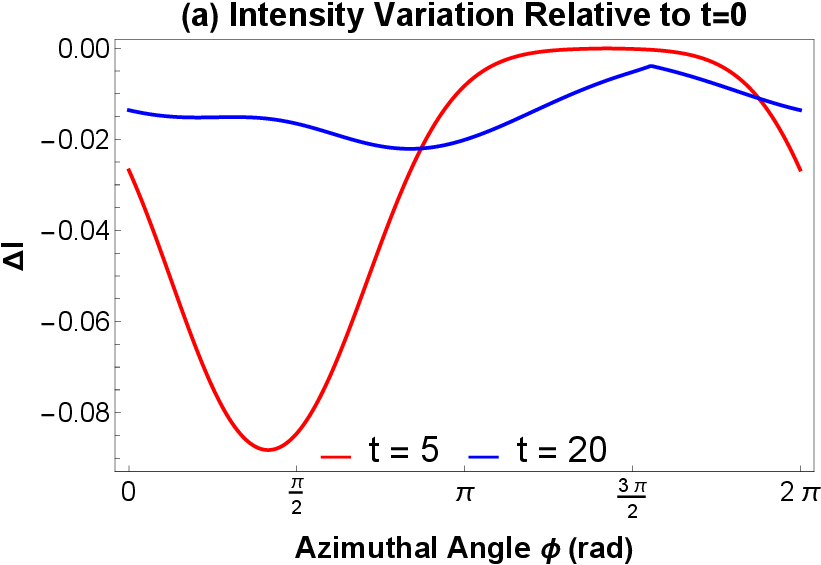}
	\includegraphics[width=4cm,height=3.5cm]{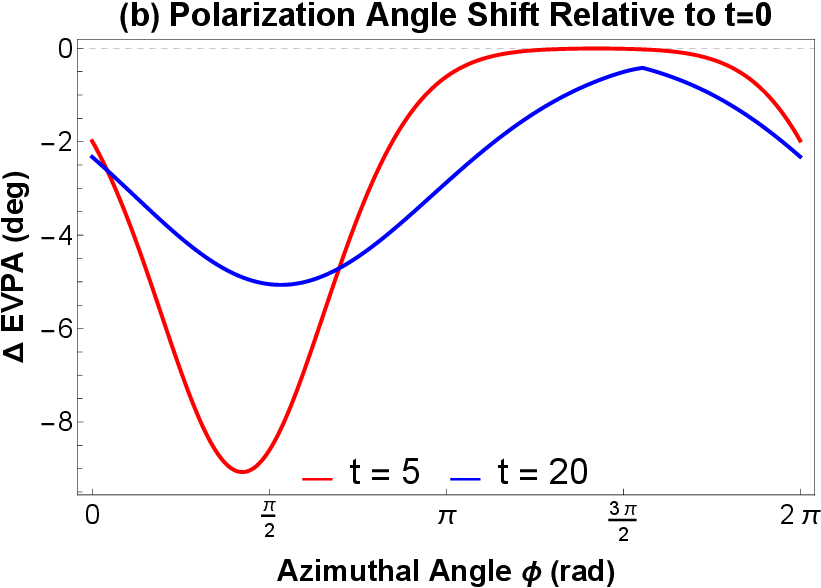}
	\includegraphics[width=4cm,height=4cm]{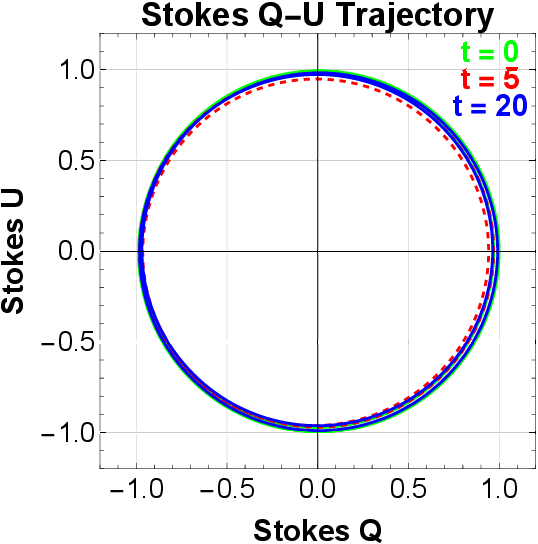}
	\includegraphics[width=4cm,height=4cm]{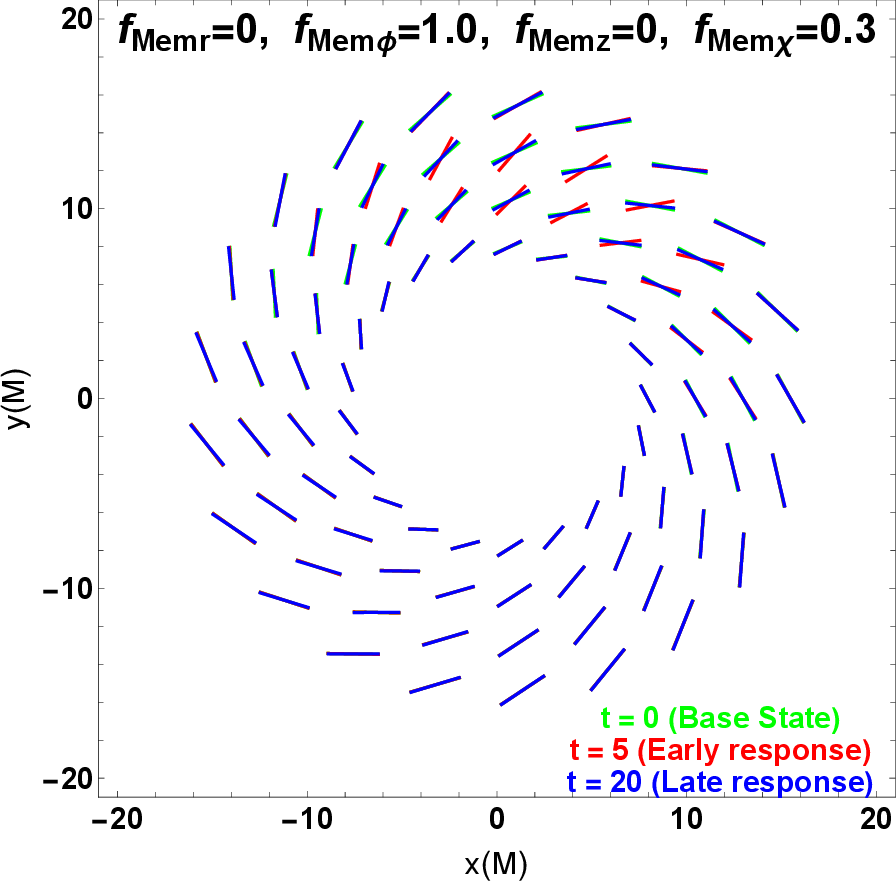}
	\includegraphics[width=4cm,height=3.5cm]{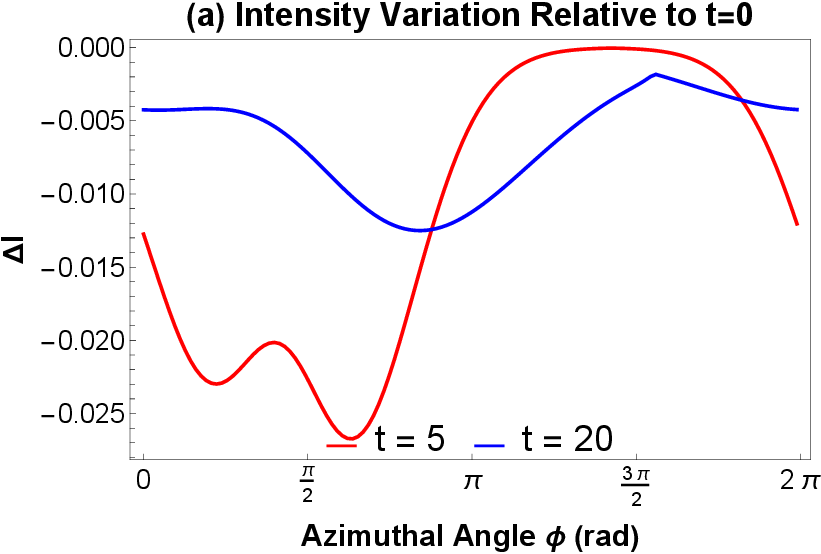}
	\includegraphics[width=4cm,height=3.5cm]{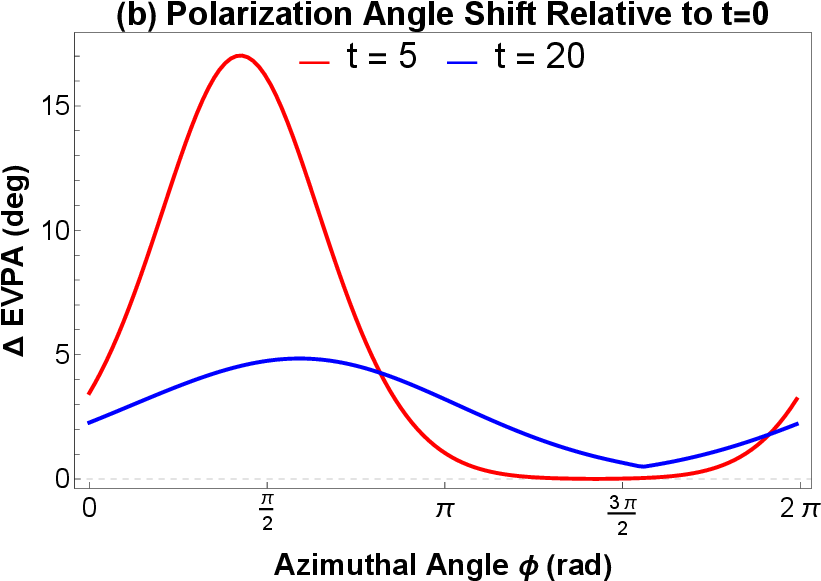}
	\includegraphics[width=4cm,height=4cm]{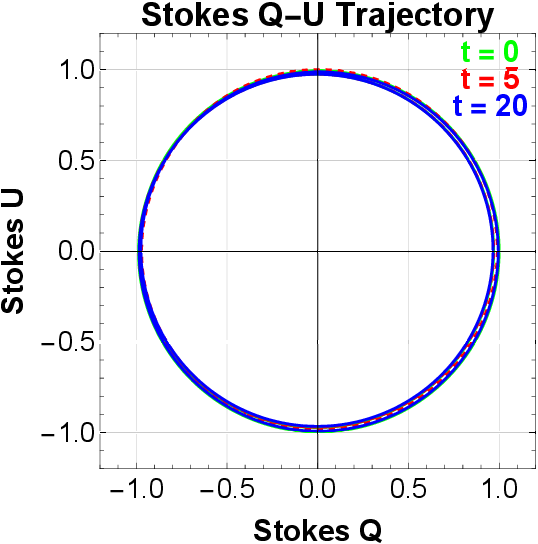}
	\includegraphics[width=4cm,height=4cm]{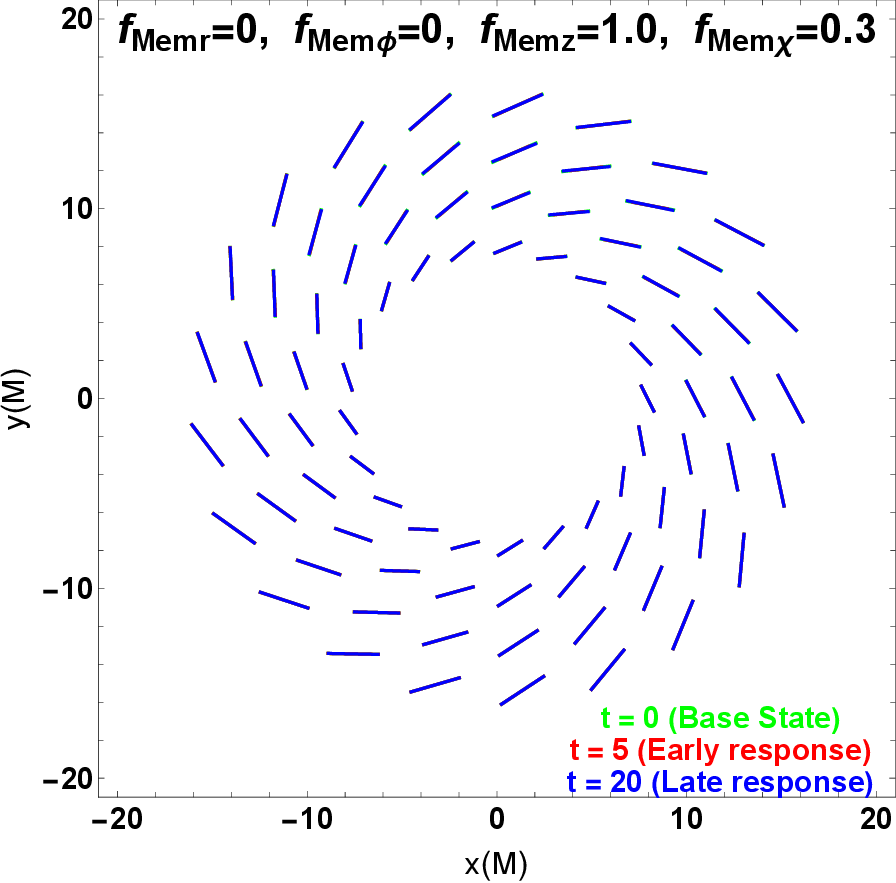}
	\includegraphics[width=4cm,height=3.5cm]{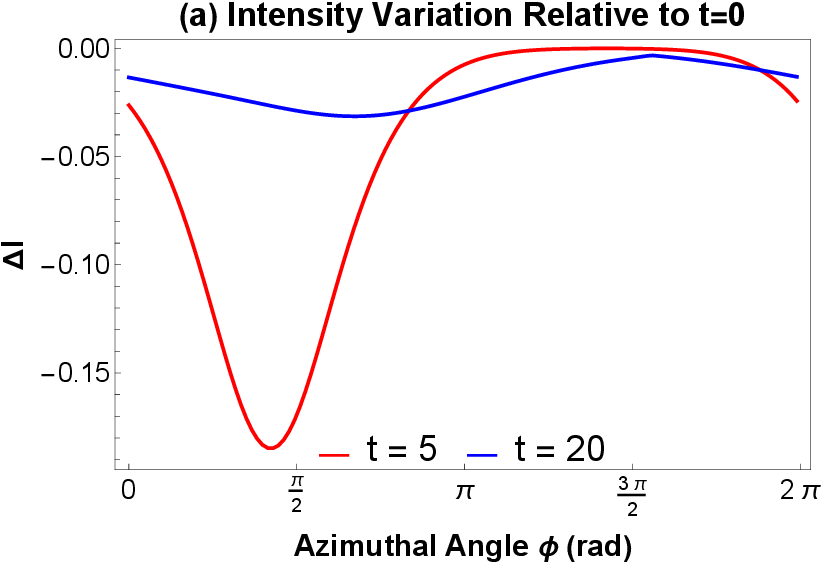}
	\includegraphics[width=4cm,height=3.5cm]{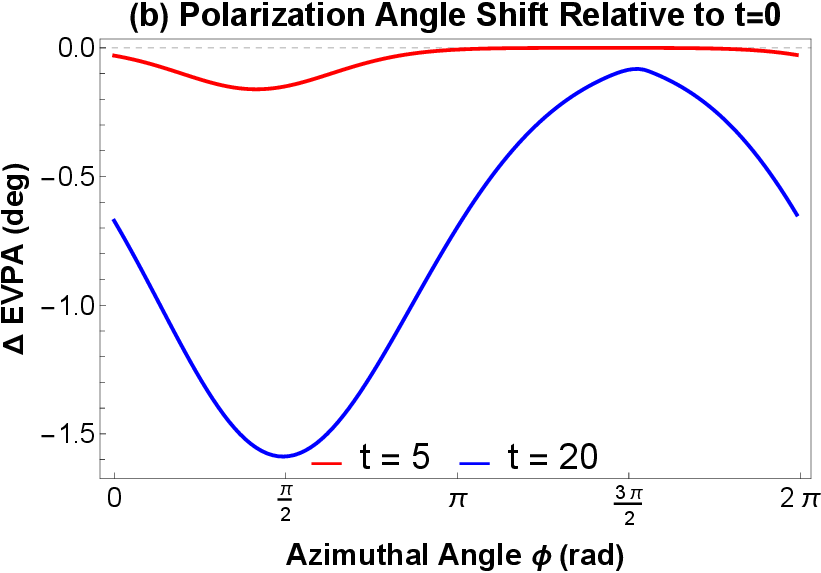}
	\includegraphics[width=4cm,height=4cm]{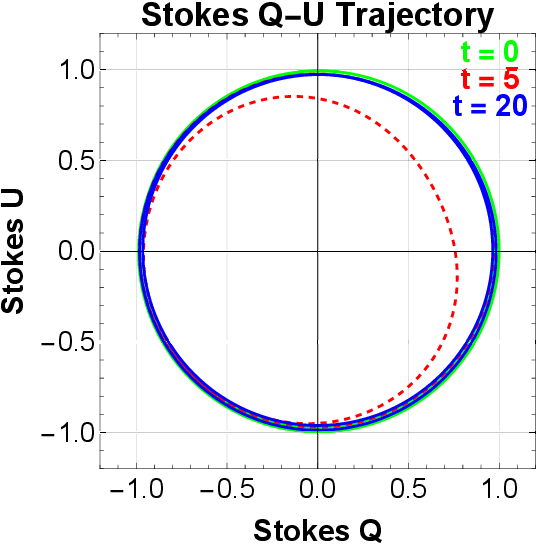}
	\caption{\label{fig:7}
		Synthetic polarization images and quantitative polarization diagnostics produced by the reconnection-bubble model. Rows correspond to cases dominated by radial (top), azimuthal (middle), and vertical (bottom) magnetic perturbations, while all other parameters are kept fixed. From left to right, the columns show representative polarization maps at \(t=0\), \(5\), and \(20\), the azimuthal polarized-intensity variation \(\Delta I\), the azimuthal EVPA residual \(\Delta{\rm EVPA}\), and the corresponding Stokes-\((Q,U)\) trajectories. The corresponding values of the memory coefficients $(f_{{\rm Mem},r}, f_{{\rm Mem},\phi}, f_{{\rm Mem},z}, f_{{\rm Mem},\chi})$ are indicated in each panel. In all panels, the remaining parameters are fixed at $A_0=0.6$, $t_{\rm inj}=1$, $r_{\rm inj}=11$, $\tau_{\rm bubble}=12$, $\tau_{\rm Mem}=20$, and $f_{{\rm Mem},\chi}=0.3$. The observer inclination is fixed at a face-on configuration.}
\end{figure*}

\subsection{Illustrative M87$^{*}$-like polarization variability from localized magnetic reconfiguration}
\label{sec:3-2}
To place the toy model in an observational context, we orient the simulated image plane so that the symmetry axis of the emitting flow is aligned with the large-scale jet direction of M87$^{*}$. Following the convention adopted in previous polarized ring studies of M87$^{*}$-like images, we rotate the image plane by a position angle of $288^\circ$ \cite{Narayan2021} and adopt a low viewing inclination, $\theta_{\rm obs}=17^\circ$, as a fiducial M87$^{*}$-like configuration. With this choice, different sectors of the simulated ring can be associated approximately with corresponding regions on the observed sky, making it possible to discuss how localized magnetic disturbances would appear in an M87$^{*}$-like horizon-scale polarization image.

Multi-epoch EHT observations in 2017, 2018, and 2021 show that the polarization morphology of M87$^{*}$ evolves significantly from year to year. In particular, the EVPA pattern is not stationary, indicating that the projected magnetic-field geometry in the emitting plasma undergoes time-dependent restructuring. Within our framework, we therefore examine whether localized reconnection-driven perturbations can naturally generate EVPA shifts with qualitatively different trends in different parts of the ring.

Figure~8 presents two illustrative realizations. In panel~(a), the perturbation is placed in a western sector of the ring near the interface between the accretion flow and the jet funnel. This region is a natural site for localized magnetic reconfiguration, since GRMHD studies indicate that the disk--funnel boundary is associated with strong magnetic shear and intermittent current-sheet activity. We therefore consider a post-reconnection response dominated by the azimuthal magnetic component, with $f_{{\rm Mem},\phi}=0.98$ and a small vertical contribution $f_{{\rm Mem},z}=0.22$, while keeping the remaining parameters fixed at $t_{\rm inj}=1$, $r_{\rm inj}=11$, $\tau_{\rm bubble}=12$, $\tau_{\rm Mem}=20$, and $f_{{\rm Mem},\chi}=0.3$. This azimuthally weighted configuration is physically motivated by the fact that, near the disk--funnel interface, the background field is strongly shaped by orbital shear, such that reconnection within a sheared current sheet can leave behind a post-reconnection geometry with a substantial toroidal imprint. As shown in Figure~8(a), the perturbation produces an outward displacement of the EVPA pattern within the affected sector of the ring. The dominant azimuthal component maintains a coherent toroidal field structure in the plane of the sky, thereby driving a systematic outward rotation of the local EVPA morphology. Meanwhile, the additional vertical component tilts part of the magnetic field toward the line of sight, reducing the effective magnetic field projected perpendicular to the photon propagation direction and consequently suppressing the local polarized synchrotron emissivity. This produces a localized weakening of the polarization intensity while preserving the overall outward EVPA excursion. The combined behavior illustrates how reconnection-driven magnetic reconfiguration can simultaneously generate coherent EVPA displacements and localized depolarization features on horizon scales.

Panel~(b) illustrates a different possibility. Here the perturbation is placed in a northeastern sector of the ring, and the post-reconnection response is taken to be dominated by the radial magnetic component, with $f_{{\rm Mem},r}=1$, while the other bubble and memory parameters are kept at the same fiducial values. A radially weighted response is physically plausible during the relaxation phase of reconnection, when newly reconnected field lines retract, straighten, or transiently stretch as the disturbed magnetic structure evolves. In such a situation, the local field geometry can acquire a stronger radial projection even though the initial instability need not have been radial in origin. In the present framework, this type of response enhances the radial projection of the magnetic field in the disturbed region and correspondingly shifts the EVPA pattern inward on the image plane. Figure~8(b) therefore illustrates another possible class of post-reconnection geometry capable of producing an inward EVPA displacement in a localized part of the ring.

To quantify the corresponding large-scale EVPA response, we compute the \(m=2\) Fourier coefficient of the complex linear polarization sampled along the emitting ring in the image plane, adopting the rotational-symmetry diagnostic introduced by Palumbo \textit{et al.}~\cite{Palumbo:2020flt} together with the corresponding thin-ring Fourier convention employed by Narayan \textit{et al.}~\cite{Narayan2021}:
\begin{equation}
	P(\varphi)
	\equiv
	Q(\varphi)+iU(\varphi)
	=
	\frac{1}{2\pi}
	\sum_{m=-\infty}^{\infty}
	\beta_m e^{im\varphi},
\end{equation}
where, with this convention, the mode coefficient is
\begin{equation}
		\beta_m
		=
		\int_0^{2\pi}
		P(\varphi)e^{-im\varphi}\,d\varphi .
\end{equation}

The coefficient \(\beta_2\) corresponds to the same \(m=2\) azimuthal polarization mode used in EHT polarimetric analyses to characterize the global twist of the resolved EVPA pattern~\cite{EventHorizonTelescope:2021bee,EventHorizonTelescope:2021srq}. In our calculation, \(P(\varphi)\) is evaluated from the polarized intensity and EVPA obtained along the sampled emitting ring, with \(\varphi\) denoting the image-plane polar angle of each ray. We focus on the phase \(\arg(\beta_2)\), which encodes the global EVPA twist relative to the local radial direction on the image plane.

Figure~9 shows the resulting evolution of \(\arg(\beta_2)\) for the two perturbations shown in Fig.~8. These scenarios represent two different post-reconnection magnetic configurations: an azimuthally dominated response in the western sector and a radially dominated response in the northeastern sector. The azimuthally dominated western-sector perturbation near the disk--funnel interface temporarily shifts \(\arg(\beta_2)\) toward less negative values, corresponding to a smaller clockwise EVPA offset relative to the local radial direction on the image plane. The phase then gradually relaxes toward its background value as the bubble response fades. By contrast, the radially dominated northeastern perturbation follows a distinct track with a smaller phase excursion. This difference arises because the contribution of a local Stokes perturbation to the global polarization pattern depends on both its polarization orientation and its position on the image ring. The sector-dependent EVPA distortions in Fig.~8 are therefore also encoded in the global \(\beta_2\)-phase response.

Taken together, the image-domain EVPA morphology and the \(\beta_2\)-phase evolution provide complementary diagnostics of localized magnetic reconfiguration: the former traces the local displacement of the polarization pattern, while the latter measures the corresponding change in the global EVPA twist. These examples show that the observed polarization response depends on both the dominant direction of post-reconnection magnetic reorganization and the location of the disturbance. Since current sheets, flux ropes, and magnetically disturbed regions arise naturally in GRMHD accretion flows, localized reconnection provides a physically motivated mechanism for producing time-dependent horizon-scale polarization variability.

\begin{figure*}[tbp]
	\centering
	\includegraphics[width=6cm,height=5.5cm]{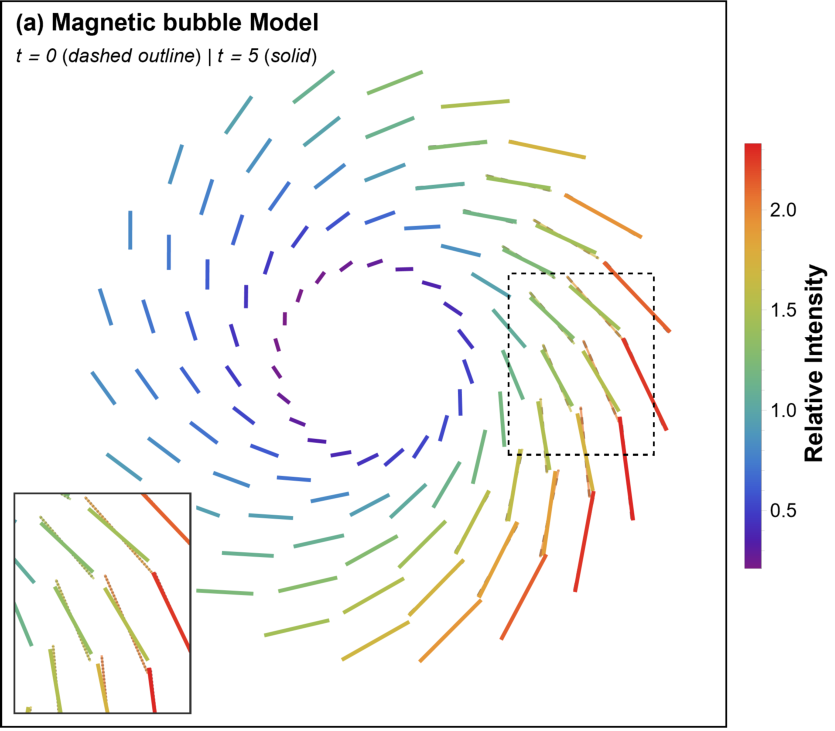}
	\includegraphics[width=6cm,height=5.5cm]{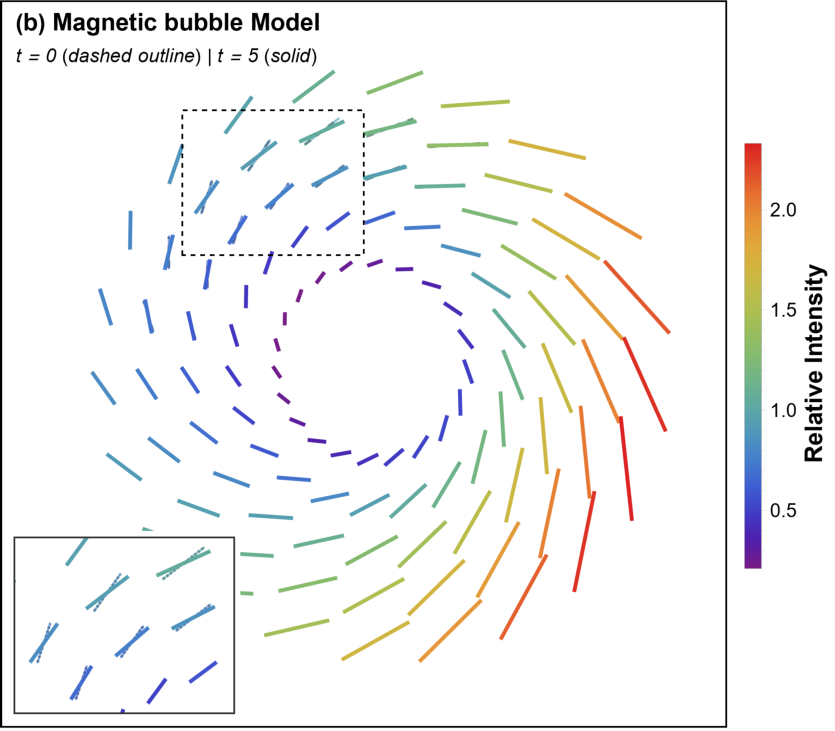}
	\caption{\label{fig:8} Illustrative M87$^{*}$-like EVPA signatures produced by localized reconnection-driven perturbations. The image plane is rotated to the observed sky orientation of M87$^{*}$. Panel (a): western-sector perturbation near the disk--funnel interface, with an azimuthally dominated response ($A_{0}=0.4$, $f_{{\rm Mem},\phi}=0.98$, $f_{{\rm Mem},z}=0.22$). Panel (b): northeastern-sector perturbation with a radially dominated response ($A_{0}=0.6$, $f_{{\rm Mem},r}=1$). In both panels, the remaining parameters are fixed at $\theta_{\rm obs}=17^\circ$, $t_{\rm inj}=1$, $r_{\rm inj}=11$, $\tau_{\rm bubble}=12$, $\tau_{\rm Mem}=20$, and $f_{{\rm Mem},\chi}=0.3$.}
\end{figure*}

\begin{figure}[tbp]
	\centering
	\includegraphics[width=7.5cm,height=5.5cm]{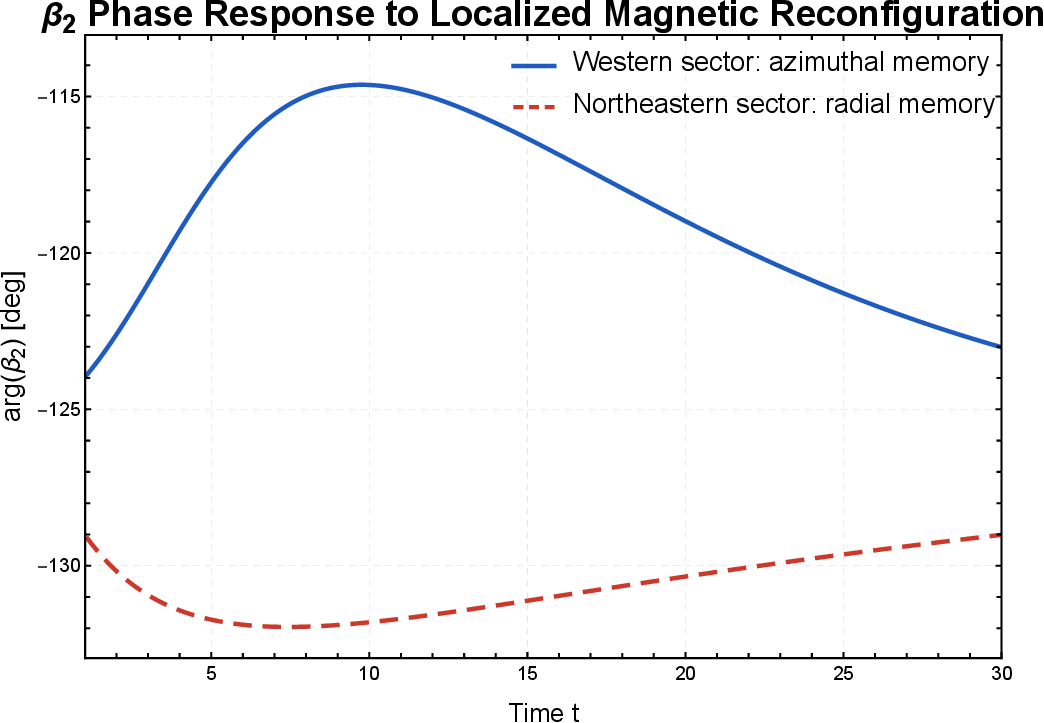}
	\caption{\label{fig:9}Time evolution of \(\arg(\beta_2)\) for the two M87\(^*\)-like reconnection-bubble cases shown in Fig.~8, with the phase analysis starting from \(t=1.1\). The blue solid and red dashed curves represent the western-sector azimuthal-memory perturbation and the northeastern-sector radial-memory perturbation, respectively.}
\end{figure}

\section{Conclusions}
\label{sec:4}
In this work, we developed a time-dependent, physically motivated toy framework to study how localized reconnection events can imprint themselves on synchrotron polarization near a black hole. Building on a polarized general relativistic ray-tracing model in Schwarzschild spacetime, we introduced a localized magnetic ``bubble'' to represent a reconnection-dominated region in which the magnetic field is temporarily depleted, its direction is reorganized, and the post-reconnection geometry relaxes over a finite time through a magnetic-memory response. We also included a conservative local kinetic response, allowing the ordered magnetic depletion to affect the polarized emission through changes in the local plasma velocity.

The model contains several key ingredients. The parameter \(A_0\) controls the residual magnetic-field amplitude inside the bubble and therefore sets the depth of the initial magnetic depletion. In this prescription, this depletion also sets the amplitude-based magnetic-energy tracer that drives the local kinetic response through \(\epsilon_{\rm kin}\) and \(\xi_B\). The timescale \(\tau_{\rm bubble}\) controls how long the depleted state persists before the field relaxes toward the background configuration. The spreading parameters \(D_i\) describe how the disturbed region broadens spatially under the action of mixing and shear in the surrounding flow. The magnetic-memory timescale \(\tau_{\rm Mem}\) characterizes the delayed relaxation of the post-reconnection field geometry, allowing the magnetic perturbation to remain dynamically relevant even after the bubble amplitude itself has started to decay. Together, these ingredients provide a controlled description of how reconnection-driven magnetic activity evolves in amplitude, geometry, and local kinematics.

Using this framework, we first examined the basic dynamical behavior of the bubble and its associated magnetic response. We showed that smaller \(A_0\) produces a deeper initial suppression of the field strength, whereas larger \(\tau_{\rm bubble}\) leads to a more prolonged recovery. Larger spreading coefficients generate a faster spatial redistribution of the perturbation, and increasing \(\tau_{\rm Mem}\) broadens the temporal profile of the memory response rather than simply rescaling its amplitude. The resulting evolution of the total magnetic-field strength demonstrates that a localized reconnection event can simultaneously produce a weakening of the field near the injection site and a growing perturbation footprint over a broader portion of the flow.

We then investigated how different post-reconnection magnetic geometries affect polarized emission when the local kinetic response is included. The results show that component-dominated perturbations are distinguishable through their combined polarized-intensity, EVPA, and Stokes-\((Q,U)\) behavior. In the examples considered here, the bubble-affected region generally shows polarized-intensity suppression because the observed polarized emission is jointly affected by ordered-field depletion, field reorientation, Doppler boosting, aberration, and pitch-angle changes. A radial-dominated response produces a moderate intensity decrease together with an inward EVPA shift. An azimuthal-dominated response gives an outward EVPA shift while retaining a milder local intensity suppression. A vertical-dominated response mainly produces the strongest polarized-intensity suppression and a comparatively weaker negative EVPA residual. These trends demonstrate that the dominant post-reconnection magnetic direction can be diagnosed through the sign and amplitude of \(\Delta{\rm EVPA}\), the strength of \(\Delta I\), and the associated Stokes-\((Q,U)\) trajectory.

To place the model in an observational context, we further considered an M87\(^*\)-like viewing geometry and illustrated how localized perturbations in different sectors of the ring can generate different classes of EVPA variability. A perturbation near the western disk--funnel boundary with an azimuthally dominated response provides one plausible way to produce outward EVPA excursions, while a perturbation in the northeastern sector with a radially dominated response illustrates a different possibility leading to inward EVPA displacements. The evolution of the \(\beta_2\) phase further shows that such localized perturbations can contribute coherently to the global EVPA twist of the ring. These examples are not intended as one-to-one reconstructions of any specific EHT epoch. Rather, they demonstrate that localized reconnection-driven magnetic restructuring can naturally generate different qualitative trends of horizon-scale EVPA variability, depending on both the location of the disturbance and the dominant direction of post-reconnection magnetic reorganization.

Overall, this work shows that localized magnetic reconnection can leave observable and physically interpretable imprints on horizon-scale polarization. The main significance of the present framework is that it provides an intermediate description between fully turbulent GRMHD simulations and direct polarimetric observables: it isolates how magnetic depletion, spatial spreading, finite-time post-reconnection memory, and local kinetic response each contribute to the evolving polarization signal. In this sense, the model offers a useful tool for interpreting time-dependent polarimetric variability and for exploring how magnetic energy release near black holes may be encoded in future horizon-scale observations.

The present framework remains phenomenological. In particular, the local kinetic response is implemented through a simplified proportional prescription that preserves the basic connection between ordered magnetic-energy release and plasma acceleration, without explicitly evolving the full energy partition among bulk motion, thermal heating, nonthermal particles, and turbulence. Future extensions may incorporate more detailed plasma thermodynamics, Kerr spacetime, finite-thickness emission regions, and polarized radiative-transfer effects.

\begin{acknowledgments}
We thank the anonymous referee for helpful comments. 
This work is supported by the National Key R\&D Program (No. 2024YFA161100), the Guangxi Science and Technology Innovation Platform Program (Leitai Action Plan, Grant No. Guike LT2600640026), Guangxi Key R\&D Program (Guangxi Funeng Action Plan, Grant No. FN2504240030), National Natural Science Foundation of China (grant No. 12133003), and the Guangxi University Outstanding Doctoral Candidate Cultivation Program Funded
Project (No. GXUODC2026058). E. W. L. is also supported by the Guangxi Talent Program (``Highland of Innovation Talents'').

\end{acknowledgments}

\nocite{*}
\bibliographystyle{unsrt}
\bibliography{apssamp}

@article{EventHorizonTelescope:2019dse,
    author = "Akiyama, Kazunori and others",
    collaboration = "Event Horizon Telescope",
    title = "{First M87 Event Horizon Telescope Results. I. The Shadow of the Supermassive Black Hole}",
    eprint = "1906.11238",
    archivePrefix = "arXiv",
    primaryClass = "astro-ph.GA",
    doi = "10.3847/2041-8213/ab0ec7",
    journal = "Astrophys. J. Lett.",
    volume = "875",
    pages = "L1",
    year = "2019"
}

@article{EventHorizonTelescope:2022wkp,
    author = "Akiyama, Kazunori and others",
    collaboration = "Event Horizon Telescope",
    title = "{First Sagittarius A* Event Horizon Telescope Results. I. The Shadow of the Supermassive Black Hole in the Center of the Milky Way}",
    eprint = "2311.08680",
    archivePrefix = "arXiv",
    primaryClass = "astro-ph.HE",
    doi = "10.3847/2041-8213/ac6674",
    journal = "Astrophys. J. Lett.",
    volume = "930",
    number = "2",
    pages = "L12",
    year = "2022"
}

@article{EventHorizonTelescope:2019pgp,
    author = "Akiyama, Kazunori and others",
    collaboration = "Event Horizon Telescope",
    title = "{First M87 Event Horizon Telescope Results. V. Physical Origin of the Asymmetric Ring}",
    eprint = "1906.11242",
    archivePrefix = "arXiv",
    primaryClass = "astro-ph.GA",
    doi = "10.3847/2041-8213/ab0f43",
    journal = "Astrophys. J. Lett.",
    volume = "875",
    number = "1",
    pages = "L5",
    year = "2019"
}

@article{EventHorizonTelescope:2019ggy,
    author = "Akiyama, Kazunori and others",
    collaboration = "Event Horizon Telescope",
    title = "{First M87 Event Horizon Telescope Results. VI. The Shadow and Mass of the Central Black Hole}",
    eprint = "1906.11243",
    archivePrefix = "arXiv",
    primaryClass = "astro-ph.GA",
    doi = "10.3847/2041-8213/ab1141",
    journal = "Astrophys. J. Lett.",
    volume = "875",
    number = "1",
    pages = "L6",
    year = "2019"
}

@article{EventHorizonTelescope:2022urf,
    author = "Akiyama, Kazunori and others",
    collaboration = "Event Horizon Telescope",
    title = "{First Sagittarius A* Event Horizon Telescope Results. V. Testing Astrophysical Models of the Galactic Center Black Hole}",
    eprint = "2311.09478",
    archivePrefix = "arXiv",
    primaryClass = "astro-ph.HE",
    reportNumber = "FERMILAB-PUB-22-419-PPD",
    doi = "10.3847/2041-8213/ac6672",
    journal = "Astrophys. J. Lett.",
    volume = "930",
    number = "2",
    pages = "L16",
    year = "2022"
}

@article{Qin:2021xvx,
    author = "Qin, Xin and Chen, Songbai and Fu, Xiangyun and Huang, Yang and Jing, Jiliang",
    title = "{Polarized image of an equatorial emitting ring around a rotating braneworld black hole}",
    eprint = "2111.10138",
    archivePrefix = "arXiv",
    primaryClass = "gr-qc",
    doi = "10.1140/epjc/s10052-025-14285-9",
    journal = "Eur. Phys. J. C",
    volume = "82",
    number = "9",
    pages = "784",
    year = "2022"
}

@article{Delijski:2022jjj,
    author = "Delijski, Valentin and Gyulchev, Galin and Nedkova, Petya and Yazadjiev, Stoytcho",
    title = "{Polarized image of equatorial emission in horizonless spacetimes: Traversable wormholes}",
    eprint = "2206.09455",
    archivePrefix = "arXiv",
    primaryClass = "gr-qc",
    doi = "10.1103/PhysRevD.106.104024",
    journal = "Phys. Rev. D",
    volume = "106",
    number = "10",
    pages = "104024",
    year = "2022"
}

@article{zhang2024polarized,
  title={Polarized images of charged particles in vortical motions around a magnetized Kerr black hole},
  author={Zhang, Zhenyu and Hou, Yehui and Hu, Zezhou and Guo, Minyong and Chen, Bin},
  journal={Journal of Cosmology and Astroparticle Physics},
  volume={2024},
  number={03},
  pages={013},
  year={2024},
  publisher={IOP Publishing}
}

@article{Zhang:2022klr,
    author = "Zhang, Zelin and Chen, Songbai and Jing, Jiliang",
    title = "{Image of Bonnor black dihole with a thin accretion disk and its polarization information}",
    eprint = "2205.13696",
    archivePrefix = "arXiv",
    primaryClass = "gr-qc",
    doi = "10.1140/epjc/s10052-022-10794-z",
    journal = "Eur. Phys. J. C",
    volume = "82",
    number = "9",
    pages = "835",
    year = "2022"
}

@article{Tlemissov2025,
  author = {Tlemissov, Abylaikhan and Toshmatov, Bobir and Kovar, Jiri},
  title = {Effect of nonlinear electrodynamics on polarization distribution around black holes},
  journal = {Physical Review D},
  volume = {111},
  pages = {064084},
  year = {2025},
  doi = {10.1103/PhysRevD.111.064084}
}

@article{huang2024coport,
  title={Coport: a new public code for polarized radiative transfer in a covariant framework},
  author={Huang, Jiewei and Zheng, Liheng and Guo, Minyong and Chen, Bin},
  journal={Journal of Cosmology and Astroparticle Physics},
  volume={2024},
  number={11},
  pages={054},
  year={2024},
  publisher={IOP Publishing}
}

@article{Aslam:2025hgl,
    author = "Aslam, Muhammad Israr and Saleem, Rabia and Yang, Chen-Yu and Zeng, Xiao-Xiong",
    title = "{Imprints of dark matter on the shadow and polarization images of a black hole illuminated by various thick disks}",
    eprint = "2512.00964",
    archivePrefix = "arXiv",
    primaryClass = "gr-qc",
    doi = "10.1016/j.jheap.2026.100551",
    journal = "JHEAp",
    volume = "51",
    pages = "100551",
    year = "2026"
}

@article{Zeng:2025kyv,
    author = "Zeng, Xiao-Xiong and Yang, Chen-Yu and Aslam, Muhammad Israr and Saleem, Rabia",
    title = "{Probing non-rotating black hole in Kalb{\textendash}Ramond gravity: imaging and polarized signatures surrounded by different thick accretion flows}",
    eprint = "2511.00586",
    archivePrefix = "arXiv",
    primaryClass = "gr-qc",
    doi = "10.1140/epjc/s10052-026-15633-z",
    journal = "Eur. Phys. J. C",
    volume = "86",
    number = "4",
    pages = "383",
    year = "2026"
}

@article{Wan:2025gbm,
    author = "Wan, Qian and Hou, Yehui and Guo, Minyong",
    title = "{Probing the scalar hair of rotating Horndeski black holes through thick disk images}",
    eprint = "2512.00917",
    archivePrefix = "arXiv",
    primaryClass = "gr-qc",
    doi = "10.1103/zjtb-3dyz",
    journal = "Phys. Rev. D",
    volume = "113",
    number = "8",
    pages = "083023",
    year = "2026"
}

@article{Hou:2024qqo,
    author = "Hou, Yehui and Huang, Jiewei and Guo, Minyong and Mizuno, Yosuke and Chen, Bin",
    title = "{Near-horizon Polarization as a Diagnostic of Black Hole Spacetime}",
    eprint = "2409.07248",
    archivePrefix = "arXiv",
    primaryClass = "gr-qc",
    reportNumber = "988 L51",
    doi = "10.3847/2041-8213/adee09",
    journal = "Astrophys. J. Lett.",
    volume = "988",
    number = "2",
    pages = "L51",
    year = "2025"
}

@article{EventHorizonTelescope:2021bee,
    author = "Akiyama, Kazunori and others",
    collaboration = "Event Horizon Telescope",
    title = "{First M87 Event Horizon Telescope Results. VII. Polarization of the Ring}",
    eprint = "2105.01169",
    archivePrefix = "arXiv",
    primaryClass = "astro-ph.HE",
    reportNumber = "FERMILAB-PUB-21-849-PPD",
    doi = "10.3847/2041-8213/abe71d",
    journal = "Astrophys. J. Lett.",
    volume = "910",
    number = "1",
    pages = "L12",
    year = "2021"
}

@article{EventHorizonTelescope:2021srq,
    author = "Akiyama, Kazunori and others",
    collaboration = "Event Horizon Telescope",
    title = "{First M87 Event Horizon Telescope Results. VIII. Magnetic Field Structure near The Event Horizon}",
    eprint = "2105.01173",
    archivePrefix = "arXiv",
    primaryClass = "astro-ph.HE",
    reportNumber = "FERMILAB-PUB-21-850-PPD",
    doi = "10.3847/2041-8213/abe4de",
    journal = "Astrophys. J. Lett.",
    volume = "910",
    number = "1",
    pages = "L13",
    year = "2021"
}

@article{EventHorizonTelescope:2024hpu,
    author = "Akiyama, Kazunori and others",
    collaboration = "Event Horizon Telescope",
    title = "{First Sagittarius A* Event Horizon Telescope Results. VII. Polarization of the Ring}",
    doi = "10.3847/2041-8213/ad2df0",
    journal = "Astrophys. J. Lett.",
    volume = "964",
    number = "2",
    pages = "L25",
    year = "2024"
}

@article{EventHorizonTelescope:2024rju,
    author = "Akiyama, Kazunori and others",
    collaboration = "Event Horizon Telescope",
    title = "{First Sagittarius A* Event Horizon Telescope Results. VIII. Physical Interpretation of the Polarized Ring}",
    doi = "10.3847/2041-8213/ad2df1",
    journal = "Astrophys. J. Lett.",
    volume = "964",
    number = "2",
    pages = "L26",
    year = "2024"
}

@article{EventHorizonTelescope:2025vum,
    author = "Akiyama, Kazunori and others",
    collaboration = "Event Horizon Telescope",
    title = "{Horizon-scale variability of M87* from 2017{\textendash}2021 EHT observations}",
    eprint = "2509.24593",
    archivePrefix = "arXiv",
    primaryClass = "astro-ph.HE",
    reportNumber = "FERMILAB-PUB-25-0735-PPD",
    doi = "10.1051/0004-6361/202555855",
    journal = "Astron. Astrophys.",
    volume = "704",
    pages = "A91",
    year = "2025"
}

@article{Chael:2021rjo,
    author = "Chael, Andrew and Johnson, Michael D. and Lupsasca, Alexandru",
    title = "{Observing the Inner Shadow of a Black Hole: A Direct View of the Event Horizon}",
    eprint = "2106.00683",
    archivePrefix = "arXiv",
    primaryClass = "astro-ph.HE",
    doi = "10.3847/1538-4357/ac09ee",
    journal = "Astrophys. J.",
    volume = "918",
    number = "1",
    pages = "6",
    year = "2021"
}

@article{Mizuno:2021esc,
    author = "Mizuno, Yosuke and Fromm, Christian M. and Younsi, Ziri and Porth, Oliver and Olivares, Hector and Rezzolla, Luciano",
    title = "{Comparison of the ion-to-electron temperature ratio prescription: GRMHD simulations with electron thermodynamics}",
    eprint = "2106.09272",
    archivePrefix = "arXiv",
    primaryClass = "astro-ph.HE",
    doi = "10.1093/mnras/stab1753",
    journal = "Mon. Not. Roy. Astron. Soc.",
    volume = "506",
    number = "1",
    pages = "741--758",
    year = "2021"
}

@article{Yuan:2022mkw,
    author = "Yuan, Feng and Wang, Haiyang and Yang, Hai",
    title = "{The Accretion Flow in M87 is Really MAD}",
    eprint = "2201.00512",
    archivePrefix = "arXiv",
    primaryClass = "astro-ph.HE",
    doi = "10.3847/1538-4357/ac4714",
    journal = "Astrophys. J.",
    volume = "924",
    number = "2",
    pages = "124",
    year = "2022"
}

@article{Yang:2024kpz,
    author = "Yang, Hai and Yuan, Feng and Li, Hui and Mizuno, Yosuke and Guo, Fan and Lu, Rusen and Ho, Luis C. and Lin, Xi and Zdziarski, Andrzej A. and Wang, Jieshuang",
    title = "{Modeling the inner part of the jet in M87: Confronting jet morphology with theory}",
    eprint = "2403.15950",
    archivePrefix = "arXiv",
    primaryClass = "astro-ph.HE",
    doi = "10.1126/sciadv.adn3544",
    journal = "Sci. Adv.",
    volume = "10",
    number = "12",
    pages = "adn3544",
    year = "2024"
}

@article{Kagan:2014hea,
    author = "Kagan, Daniel and Sironi, Lorenzo and Cerutti, Benoit and Giannios, Dimitrios",
    title = "{Relativistic magnetic reconnection in pair plasmas and its astrophysical applications}",
    eprint = "1412.2451",
    archivePrefix = "arXiv",
    primaryClass = "astro-ph.HE",
    doi = "10.1007/s11214-014-0132-9",
    journal = "Space Sci. Rev.",
    volume = "191",
    number = "1-4",
    pages = "545--573",
    year = "2015"
}

@article{Ball:2018,
    author = {Ball, David and {\"O}zel, Feryal and Psaltis, Dimitrios and Chan, Chi-Kwan and Sironi, Lorenzo},
    title = "{The Properties of Reconnection Current Sheets in GRMHD Simulations of Radiatively Inefficient Accretion Flows}",
    eprint = "1705.06293",
    archivePrefix = "arXiv",
    primaryClass = "astro-ph.HE",
    doi = "10.3847/1538-4357/aaa42f",
    journal = "Astrophys. J.",
    volume = "853",
    number = "2",
    pages = "184",
    year = "2018"
}

@article{Nalewajko:2024qtc,
    author = "Nalewajko, Krzysztof and Kapusta, Mateusz and Janiuk, Agnieszka",
    title = "{Chaotic magnetic disconnections trigger flux eruptions in accretion flows channeled onto magnetically saturated Kerr black holes}",
    eprint = "2410.08280",
    archivePrefix = "arXiv",
    primaryClass = "astro-ph.HE",
    doi = "10.1051/0004-6361/202450490",
    journal = "Astron. Astrophys.",
    volume = "692",
    pages = "A37",
    year = "2024"
}

@article{Porth:2020txf,
    author = "Porth, O. and Mizuno, Y. and Younsi, Z. and Fromm, C. M.",
    title = "{Flares in the Galactic Centre {\textendash} I. Orbiting flux tubes in magnetically arrested black hole accretion discs}",
    eprint = "2006.03658",
    archivePrefix = "arXiv",
    primaryClass = "astro-ph.HE",
    doi = "10.1093/mnras/stab163",
    journal = "Mon. Not. Roy. Astron. Soc.",
    volume = "502",
    number = "2",
    pages = "2023--2032",
    year = "2021"
}

@article{Najafi-Ziyazi:2023oil,
    author = "Najafi-Ziyazi, Mahdi and Davelaar, Jordy and Mizuno, Yosuke and Porth, Oliver",
    title = "{Flares in the Galactic centre {\textendash} II. Polarization signatures of flares at mm-wavelengths}",
    eprint = "2308.16740",
    archivePrefix = "arXiv",
    primaryClass = "astro-ph.HE",
    doi = "10.1093/mnras/stae1343",
    journal = "Mon. Not. Roy. Astron. Soc.",
    volume = "531",
    number = "4",
    pages = "3961--3972",
    year = "2024"
}

@article{Jiang:2024gtk,
    author = "Jiang, Hong-Xuan and Mizuno, Yosuke and Dihingia, Indu K. and Nathanail, Antonios and Younsi, Ziri and Fromm, Christian M.",
    title = "{Dynamics and emission properties of flux ropes from two-temperature GRMHD simulations with multiple magnetic loops}",
    eprint = "2404.03237",
    archivePrefix = "arXiv",
    primaryClass = "astro-ph.HE",
    doi = "10.1051/0004-6361/202449681",
    journal = "Astron. Astrophys.",
    volume = "688",
    pages = "A82",
    year = "2024"
}

@article{Jiang:2025huk,
    author = "Jiang, Hong-Xuan and Mizuno, Yosuke and Dihingia, Indu K. and Yuan, Feng and Lin, Xi and Fromm, Christian M. and Nathanail, Antonios and Younsi, Ziri",
    title = "{The Physical Origin and Time Lag of Multifrequency Flares from Sgr A*}",
    eprint = "2507.12789",
    archivePrefix = "arXiv",
    primaryClass = "astro-ph.HE",
    doi = "10.3847/1538-4357/adf1e5",
    journal = "Astrophys. J.",
    volume = "990",
    number = "1",
    pages = "81",
    year = "2025"
}

@article{Nathanail:2020wap,
    author = "Nathanail, Antonios and Fromm, Christian M. and Porth, Oliver and Olivares, Hector and Younsi, Ziri and Mizuno, Yosuke and Rezzolla, Luciano",
    title = "{Plasmoid formation in global GRMHD simulations and AGN flares}",
    eprint = "2002.01777",
    archivePrefix = "arXiv",
    primaryClass = "astro-ph.HE",
    doi = "10.1093/mnras/staa1165",
    journal = "Mon. Not. Roy. Astron. Soc.",
    volume = "495",
    number = "2",
    pages = "1549--1565",
    year = "2020"
}

@article{Ripperda:2020bpz,
    author = "Ripperda, Bart and Bacchini, Fabio and Philippov, Alexander",
    title = "{Magnetic Reconnection and Hot Spot Formation in Black Hole Accretion Disks}",
    eprint = "2003.04330",
    archivePrefix = "arXiv",
    primaryClass = "astro-ph.HE",
    doi = "10.3847/1538-4357/ababab",
    journal = "Astrophys. J.",
    volume = "900",
    number = "2",
    pages = "100",
    year = "2020"
}

@article{GrigorianDexter:2024,
    author = "Grigorian, Arpiar Avetis and Dexter, Jason",
    title = "{The relationship between simulated sub-millimeter and near-infrared images of Sagittarius A* from a magnetically arrested black hole accretion flow}",
    eprint = "2404.10982",
    archivePrefix = "arXiv",
    primaryClass = "astro-ph.HE",
    doi = "10.1093/mnras/stae934",
    journal = "Mon. Not. Roy. Astron. Soc.",
    volume = "530",
    number = "2",
    pages = "1563--1579",
    year = "2024"
}

@article{LinYuan:2024,
    author = "Xi, Lin and Feng, Yuan",
    title = "{Revisiting flares in Sagittarius A* based on general relativistic magnetohydrodynamic numerical simulations of black hole accretion}",
    eprint = "2405.17408",
    archivePrefix = "arXiv",
    primaryClass = "astro-ph.HE",
    doi = "10.1093/mnras/stae1357",
    journal = "Mon. Not. Roy. Astron. Soc.",
    volume = "531",
    number = "3",
    pages = "3136--3150",
    year = "2024"
}

@ARTICLE{Santhiya:2025ouh,
       author = {{Santhiya}, P.~S. and {Bhat}, Pallavi and {Kumar}, Prayush and {Mondal}, Tushar and {Dihingia}, Indu K.},
        title = "{Dynamo and Jet interconnections in GRMHD simulations of black hole accretion disks}",
      journal = {arXiv e-prints},
         year = 2025,
        month = dec,
          eid = {arXiv:2512.02129},
        pages = {arXiv:2512.02129},
          doi = {10.48550/arXiv.2512.02129},
archivePrefix = {arXiv},
       eprint = {2512.02129},
 primaryClass = {astro-ph.HE},
       adsurl = {https://ui.adsabs.harvard.edu/abs/2025arXiv251202129S}
}

@article{Narayan2021,
  author = {Narayan, Ramesh and Palumbo, Daniel C. M. and Johnson, Michael D. and Gelles, Zachary and Himwich, Elizabeth and Chang, Dominic O. and Ricarte, Angelo and Dexter, Jason and Gammie, Charles F. and Chael, Andrew A. and {The Event Horizon Telescope Collaboration}},
  title = {The Polarized Image of a Synchrotron-emitting Ring of Gas Orbiting a Black Hole},
  journal = {The Astrophysical Journal},
  volume = {912},
  pages = {35},
  year = {2021},
  doi = {10.3847/1538-4357/abf117}
}

@article{Gelles2021,
  title={Polarized image of equatorial emission in the Kerr geometry},
  author={Gelles, Zachary and Himwich, Elizabeth and Johnson, Michael D and Palumbo, Daniel CM},
  journal={Physical Review D},
  volume={104},
  number={4},
  pages={044060},
  year={2021},
  publisher={APS}
}

@article{Sironi:2025kgn,
    author = "Sironi, Lorenzo and Uzdensky, Dmitri A. and Giannios, Dimitrios",
    title = "{Relativistic Magnetic Reconnection in Astrophysical Plasmas: A Powerful Mechanism of Nonthermal Emission}",
    eprint = "2506.02101",
    archivePrefix = "arXiv",
    primaryClass = "astro-ph.HE",
    doi = "10.1146/annurev-astro-020325-115713",
    journal = "Ann. Rev. Astron. Astrophys.",
    volume = "63",
    number = "1",
    pages = "127--178",
    year = "2025"
}

@article{Palumbo:2020flt,
    author = "Palumbo, Daniel C. M. and Wong, George N. and Prather, Ben S.",
    title = "{Discriminating Accretion States via Rotational Symmetry in Simulated Polarimetric Images of M87}",
    eprint = "2004.01751",
    archivePrefix = "arXiv",
    primaryClass = "astro-ph.HE",
    doi = "10.3847/1538-4357/ab86ac",
    journal = "Astrophys. J.",
    volume = "894",
    number = "2",
    pages = "156",
    year = "2020"
}

\end{document}